\documentclass[twocolumn]{aastex7}
\usepackage[caption=false]{subfig}
\usepackage{mathtools,amssymb}
\usepackage{aligned-overset}

\begin{document}

\newcommand{\be}{\begin{equation}}
\newcommand{\ee}{\end{equation}}

\title{A Unified Model for Shock Interaction and $\gamma$-Ray Emission in Classical Novae}

\author[0000-0002-6679-0012]{Rebecca Diesing}
\affil{School of Natural Sciences, Institute for Advanced Study, Princeton, NJ 08540, USA}
\affil{Department of Physics and Columbia Astrophysics Laboratory, Columbia University, New York, NY 10027, USA}
\email{rrdiesing@ias.edu}

\author[0000-0002-4670-7509]{Brian D.~Metzger}
\affil{Department of Physics and Columbia Astrophysics Laboratory, Columbia University, New York, NY 10027, USA}
\affil{Center for Computational Astrophysics, Flatiron Institute, 162 5th Ave, New York, NY 10010, USA} 
\email{bdm2129@columbia.edu}

\begin{abstract}
We present a parameterized (``toy'') model for shock interaction and $\gamma$-ray emission in classical novae, in which a white dwarf envelope of mass $M_{\rm env}$ is removed over a timescale $\tau$ (proportional to the nova speed class, $t_{2}$) in an outflow that accelerates on the same timescale to a terminal speed $v_{\rm f}$. Particle acceleration occurs at the reverse shock generated when the outflow collides with a thin, dense shell of slower material released earlier. Accelerated protons are then advected into the shell, where for typical ${ M_{\rm env}, \tau, \text{and } v_{\rm f}}$ they radiate in the calorimetric limit, consistent with correlated optical and $\gamma$-ray emission seen in well-sampled novae. The maximum proton energy, set by a Hillas-like argument, scales with the thickness of the hot post-shock region. Recent work shows turbulent mixing of hot post-shock gas with cooler dense gas may limit this thickness to $\lesssim 10^{-4}$ of the shock radius, explaining low X-ray luminosities. Using this empirically motivated thickness, and assuming efficient magnetic amplification, we predict maximum proton energies $E_{\rm max} \sim 10$ GeV, consistent with $\gamma$-ray spectra of \emph{Fermi}-detected novae near optical peak ($\sim \tau$). However, as the shock and post-shock layer expand, $E_{\rm max}$ can grow to $\gtrsim 10$ TeV on timescales of a few $\tau$, enabling potential detection by atmospheric Cherenkov telescopes. We encourage TeV follow-up of {\it Fermi}-detected novae weeks to months after the optical/GeV peak and quantify the most promising events.
\end{abstract}

\section{Introduction}

\begin{figure*}
    \centering
    \includegraphics[width=\linewidth]{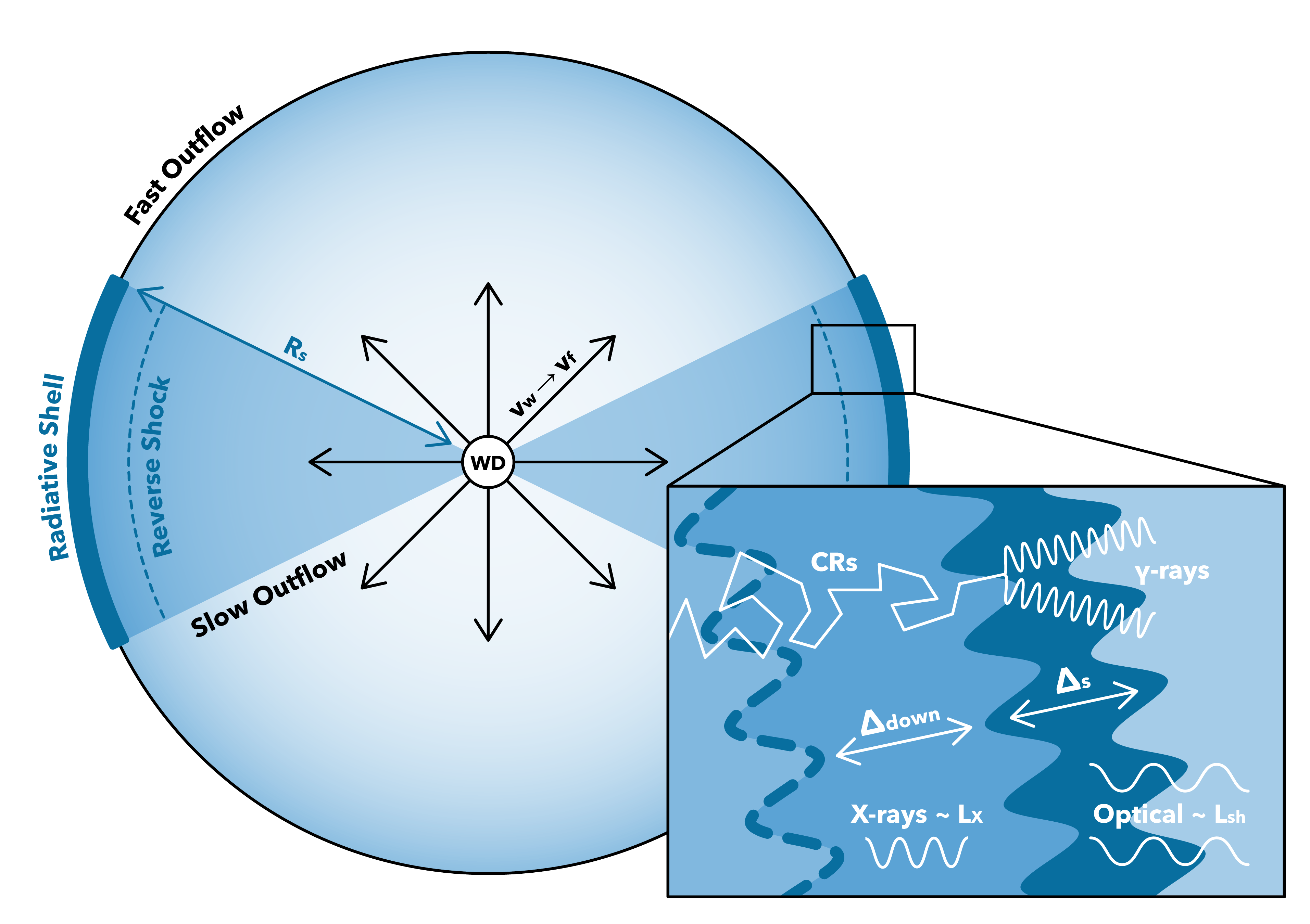}
    \caption{Illustration of the toy model for shock interaction in classical novae. A fast, spherical wind collides with slower material released earlier in the eruption, mediated by a reverse shock. The slow early ejecta is concentrated in the binary equatorial plane, subtending a fractional solid angle $f_{\Omega} \gtrsim 0.1$. The swept up gas resides in a dense, highly corrugated radiative shell, which reprocesses most of the shock power into thermal infrared/optical emission. Protons accelerated at the shock are quickly advected into this shell where they produce $\gamma$-ray emission via proton-proton collisions (inset diagram). Although the shock is radiative, its X-ray luminosity ($L_{\rm X}$) is greatly suppressed compared to the total shock power ($L_{\rm sh}$, which is comparable to the optical luminosity) due to efficient turbulent mixing between the host post-shock gas and the cool shell.  The observed ratio $f_{\rm X} = L_{\rm X}/L_{\rm sh} \sim 10^{-4}$ constrains the radial thickness of the hot postshock region, $\Delta_{\rm down}$, within which non-thermal ion acceleration occurs. As this region is extremely thin compared to either the normal laminar radiative cooling length ($\Delta_{\rm rad}$; Eq.~\eqref{eq:Deltarad}) or the overall radius of the shell ($R_{\rm s}$), $\Delta_{\rm down}$ constrains the maximum proton energy.}
    \label{fig:diagram}
\end{figure*}
Classical novae are outbursts on the surfaces of white dwarfs powered by nuclear burning of hydrogen-rich material accreted from a stellar binary companion \citep{Gallagher&Starrfield78,Wolf+13}.  They reach peak visual luminosities $\sim 10^{4}L_{\odot}$ approaching or exceeding the white dwarf Eddington luminosity and eject large quantities of mass $\sim 10^{-6}-10^{-4}M_{\odot}$ at high velocities $\sim 10^{3}$ km s$^{-1}$ \citep[e.g.,][]{Chomiuk+21ARAA}.  Although long thought to be powered directly by energy released from nuclear burning, a variety of evidence suggests that {\it shocks} play a key role in powering nova emission across the electromagnetic spectrum.  This evidence includes multiple complex velocity components in the optical spectra indicative of outflow interaction \citep[e.g.,][]{Williams+08,Aydi+20b}, $\gtrsim$ keV thermal X-ray emission from hot shocked gas weeks to months after the outburst \citep[e.g.,][]{Orio+01,Mukai+08}, and early maxima in the radio light curves with brightness temperatures in excess of those expected from $\sim 10^{4}$ K photo-ionized ejecta \citep[e.g.,][]{Chomiuk+14,Weston+16a,Chomiuk+21}.  

The most striking indicator of shocks is the discovery by {\it Fermi} LAT of $\sim 0.1-100$ GeV $\gamma$-ray emission, observed within a few days of the optical peak and lasting weeks \citep{Ackermann+14}.  The first $\gamma$-ray nova occurred in the symbiotic binary V407 Cyg \citep{Abdo+10}, suggesting that the shocks arose from the collision between the nova outflow and the dense wind of the companion giant \citep[e.g.,][]{Martin&Dubus13}.  However, $\gamma$-rays have now been detected from over 20 classical novae with main sequence stellar companions \citep{Cheung+16,Franckowiak+18,Craig+25}.  This indicates that the outflow from the white dwarf runs into similarly dense gas even in systems not embedded in a giant's wind.  This external medium likely represents slower ejecta from earlier in the outburst (i.e., the shocks are ``internal"; \citealt{Friedjung87,Metzger+14,Martin+18,hachisu+22}).

One physical picture, consistent with both optical/infrared \citep[e.g.,][]{Schaefer+14,Aydi+25} and radio imaging \citep[e.g.,][]{Chomiuk+14}, and the evolution of optical spectral lines \citep[e.g.,][]{Ribeiro+13,Shore+13,Aydi+20b}, is that the thermonuclear runaway is first accompanied by a slow ejection of mass with an equatorially-focused geometry, the shape of which is likely influenced by the gravity of the binary stellar companion \citep[e.g.,][]{Livio+90,Lloyd+97,Pejcha+16b,Pejcha+16a}.  This slow outflow is then followed by a second ejection or a continuous radiation pressure-driven wind \citep[e.g.,][]{Bath&Shaviv76,Shen&Quataert22} with a higher velocity and more spherical geometry.  The subsequent collision between the fast and slow components produces strong shocks in the ejecta, concentrated in the equatorial plane.  The fast component continues to expand freely along the polar direction, generating a bipolar morphology consistent with observations (Figure \ref{fig:diagram}).  Recent work characterizing densely time-sampled optical spectra of a large sample of novae ($\sim 12$ events) reveals strong evidence for these evolutionary phases across novae of all speed classes \citep{Aydi+20b}.  Although present samples are flux-limited in the $\gamma$-ray band due to the limited sensitivity of {\it Fermi} LAT, they are also consistent with internal shock interaction being ubiquitous \citep{Chomiuk+21ARAA}. 

Shocks in novae accelerate non-thermal ions and electrons (cosmic rays) via diffusive shock acceleration \citep[]{fermi54, krymskii77, axford+77p, bell78a, blandford+78}, in which charged particles scatter off magnetic field perturbations, resulting in diffusion across the shock and an energy gain with each crossing. This mechanism produces power-law distributions of particles with a maximum energy set by requiring that the acceleration time (which is comparable to the diffusion time) be less than the age of the system \citep{drury83} and the diffusion length be smaller than the size of the acceleration region \citep{hillas84}. Both criteria depend sensitively on the strength of the local magnetic field, which may be amplified by the propagation of the cosmic rays themselves \citep[e.g.,][]{Bell04} or other instabilities driving turbulence as described below.  

Cosmic rays accelerated at nova shocks produce $\gamma$-ray emission chiefly via the decay of neutral pions following inelastic collisions between ions and ambient gas (e.g., \citealt{Vurm&Metzger18,Martin+18}). 
The high densities behind nova shocks (which are required to reproduce the observed $\gamma$-ray fluxes) cause non-thermal protons to lose their energy rapidly compared to the outflow expansion time. Shocks in novae therefore serve as ``cosmic ray calorimeters" \citep{Metzger+15} which reflect changes in particle acceleration in real-time and provide new probes of open issues, such as the sources of cosmic rays and the amplification of magnetic fields, complementary to those in other astrophysical environments \citep[e.g.,][]{Metzger+15,Martin+18,Fang+20}. 

The high densities behind nova shocks also result in novel effects related to the influence of radiative cooling on the dynamical and emission properties of the system, i.e. the shocks are ``radiative" even from early times.  Rapid cooling leads to complex multi-dimensional behavior at and behind the shock-front driven by a combination of thermal, thin-shell, and Rayleigh-Taylor instabilities \citep{Chevalier&Imamura82,Vishniac94}.   The dense cool shell swept up by the shocks efficiently absorbs the shock's UV/X-ray radiation and reprocesses most of its power into infrared/optical wavelengths \citep{Metzger+14}, similar to interacting supernovae \citep{chevalier+94}.  The high observed ratio of $\gamma$-ray to optical luminosity implies that a significant fraction of the nova's optical light curve is reprocessed shock emission rather than direct emission from the white dwarf \citep{Metzger+15,Li+17}.  The clumpy, dense radiatively cooled shell of shocked gas also provides a shielded environment for forming molecules and dust \citep{Derdzinski+17}, another well-observed but poorly understood phenomenon (e.g.,~\citealt{Evans&Gehrz12,Shore+18,Finzell+18,Chong+25}).

While abundant evidence supports the presence of powerful shocks in novae, their (absorption-corrected) X-ray luminosities are several orders of magnitude lower than expected from their $\gamma$-ray luminosities when contemporaneous observations are available (\citealt{Nelson+19,Sokolovsky+20,Sokolovsky+22}). This unexpected suppression of the shock's X-ray thermal emission has recently been attributed to turbulent mixing between the hot immediate post-shock gas and the cool dense shell of swept up ejecta \citep{Metzger+25}. Such mixing between hot and cold gas at the reverse shock is consistent with the charge-exchange recombination line signatures recently observed in the X-ray spectra of novae \citep{Mitrani+24,Mitrani+25}.  The low observed X-ray luminosities, corresponding to a fraction $\lesssim 10^{-4}$ of the shock power, imply a commensurate reduction in the thickness of the X-ray layer relative to that of laminar models \citep{Metzger+14}.\footnote{{\it Non-thermal} hard X-ray emission from primary or secondary electrons is also strongly suppressed in the dense shell as a result of strong Coulomb losses to the thermal plasma \citep{Vurm&Metzger18}, consistent with the lack of a hard power-law component in X-ray spectra of novae taken by {\it NuSTAR}  (e.g., \citealt{Nelson+19,Sokolovsky+20}).} This effectively places the dense cool shell much closer to the shock than naively predicted from the radiative cooling length of the hot gas. As we shall describe, this realization has major implications for particle acceleration and $\gamma$-ray emission, since it reduces the width of the particle acceleration region and hence the maximum proton energy significantly compared to previous estimates (e.g., \citealt{Metzger+16}).

In this paper we consolidate the wealth of multi-wavelength observational constraints on classical nova evolution into a parameterized model for their shock interaction, particle acceleration, and $\gamma$-ray emission.  We introduce the model in Section \ref{sec:model}, in parallel with analytic estimates to help guide the discussion and motivate the parameter choices.  In Section \ref{sec:results} we describe implications of our results for the observed $\gamma$-ray emission from classical novae at both GeV and TeV energies.  As we will demonstrate, a subset of \emph{Fermi}-detectable novae at GeV energies may also be detectable at TeV energies with current-generation imaging atmospheric Cherenkov telescopes (IACTs).  In Section \ref{sec:discussion} we discuss additional implications of our results and in Section \ref{sec:conclusion} we summarize and conclude.




\section{Model}
\label{sec:model}

Herein we describe our model for shock interaction in classical nova outflows and associated $\gamma$-ray emission.  The parameters of the model and their fiducial values are summarized in Table \ref{tab:modelparams}, and will be described in detail as we proceed.

\begin{deluxetable}{ccc}
\tablecaption{Model Parameters\label{tab:modelparams}}
\tablewidth{700pt}
\tabletypesize{\scriptsize}
\tablehead{
\colhead{Symbol} & \colhead{Description} & 
\colhead{Fiducial Value (Range)} 
} 
\startdata
$M_{\rm env}$ & Envelope mass & $10^{-4}\,\,(10^{-6}-10^{-4})M_{\odot}$ \\
$\tau \approx 0.5 t_{2}$ & Envelope removal time & 20\,\,($10-75$) {\rm d} \\
$v_{\rm f} \approx 2v_{2}$ & Final wind speed & $6000\,\,(2000-10^4$) km s$^{-1}$ \\
$f_{\rm X} \equiv L_{\rm X}/L_{\rm sh}$ & X-ray efficiency & $5\times 10^{-5}\,\,(10^{-4}-10^{-5})$ \\ 
$\Delta_{\rm s}/R_{\rm s}$ &  Cool shell thickness & $10^{-2}\,\,(10^{-3}-10^{-2})$  \\
$f_{\rm \Omega} \equiv \Delta \Omega/4\pi$ & Slow outflow covering fraction & $0.3\,\,(0.1-1)$ \\ 
$\xi_{\rm CR}$ & Cosmic ray acceleration efficiency & $0.03\,\,(0.01-0.1)$ \\
$\xi_{\rm B}$ & B-field amplification efficiency & $0.01\,\,(10^{-3}-0.1)$ \\
\enddata
\end{deluxetable}

\subsection{Hydrodynamics} \label{subsec:hydro}

Consider a phenomenological model in which the white dwarf envelope of mass $M_{\rm env} \approx 10^{-6}-10^{-4}M_{\odot}$ is removed on a timescale $\tau$.  As the mechanisms of mass removal are still debated (e.g., \citealt{Chomiuk+21}), we assume a somewhat ad hoc time-dependence for the mass-loss rate of the nova outflow (``wind''):
\be
\dot{M}_{\rm w}(t) = \frac{M_{\rm env}}{\tau}e^{-t/\tau}.
\label{eq:Mdotw}
\ee 
Likewise, we assume that the velocity of the wind increases over the same timescale $\tau$,
\be
v_{\rm w}(t) = \left[1-e^{-t/\tau}\right]v_{\rm f},
\label{eq:vw}
\ee
to some ``final'', or ``fast'', value $v_{\rm f}$ of typically several thousand km s$^{-1}$.    

The mass loss parameters can be roughly mapped into nova observables.  If the nova light-curve decays in proportion to $\dot{M}_{\rm w}$, then $\tau$ is related to the nova speed class according to $\tau \approx t_{2}/2$, where $t_{\rm 2}$ is the time for the optical light-curve to decay by two magnitudes ($t_{2} \lesssim 25$ d is a fast nova, while $t_{2} \gtrsim 150$ d is a very slow nova).  Likewise, as we discuss below (Eq.~\eqref{eq:vs}), the final wind velocity $v_{\rm f}$ is roughly twice the maximum speed achieved by the shell of the cumulative ejecta. We interpret the latter as the ``fast'' component inferred spectroscopically by \citet[their ``$v_{\rm 2}$'']{Aydi+20b}, in which case $v_{\rm f} \approx 2v_{2}$. 

We assume for simplicity a spherical outflow geometry in our discussion of the ejecta dynamics throughout this section, even though the earliest phases of the nova outflow may be concentrated in the binary plane (Fig.~\ref{fig:diagram}). As we proceed, we will outline places where corrections for the potential non-spherical shock geometry enters our calculation, particularly in predicting the $\gamma$-ray emission.  

The nova outflow feeds a thin cool shell of gas, containing the earlier slow ejecta, whose mass therefore grows as
\be
M_{\rm s}(t) = \int_0^{t} \dot{M}_{\rm w} dt' = \left[1-e^{-t/\tau}\right] M_{\rm env}.
\label{eq:Ms}
\ee
The shell has a velocity $v_{\rm s}$, which in general will be smaller than $v_{\rm w}$.  The wind interacts with and adds mass and momentum to the shell through a reverse shock. The shock is assumed to be radiative, if not directly by X-rays from the immediate post-shock gas, due to turbulent mixing with and radiation from cooler gas present in the shell \citep{Metzger+25}.  This interaction is momentum conserving, such that 
\be
\frac{d}{dt}\left(M_{\rm s}v_{\rm s}\right) = \dot{M}_{\rm w}v_{\rm w}.
\ee
Under the assumption that $M_{\rm s}v_{\rm s} \rightarrow 0$ as $t \rightarrow 0$, the solution to this equation is given by
\be
v_{\rm s}(t) = \left[1-e^{-t/\tau}\right]\frac{v_{\rm f}}{2} = \frac{v_{\rm w}}{2},
\label{eq:vs}
\ee
i.e., the shell expands at half the instantaneous wind speed.  Integrating this expression gives the radius of both the shock and the shell,
\be
R_{\rm s}(t) = \frac{v_{\rm f}}{2}\left[t-\tau + \tau e^{-t/\tau}\right] \underset{t \gg \tau}\approx \frac{1}{2}v_{\rm f}(t-\tau),
\label{eq:Rs}
\ee
where in the final line, and in similar equalities below, we assume $t \gg \tau.$  The shock velocity is given by
\be
v_{\rm sh}(t) = v_{\rm w} - v_{\rm s} = \frac{v_{\rm w}}{2} = v_{\rm s},
\label{eq:vsh}
\ee
while the total kinetic power dissipated at the shock equals
\begin{equation}
\begin{split}
   L_{\rm sh}(t) & = \frac{1}{2}\dot{M}_{\rm w}\left[v_{\rm w}^{2}-v_{\rm s}^{2}\right] = \frac{3}{8}\dot{M}_{\rm w}v_{\rm w}^{2}  \\ & =  \frac{3}{8}\frac{M_{\rm env}v_{\rm f}^{2}}{\tau}e^{-t/\tau}\left[1-e^{-t/\tau}\right]^{2}. 
   \label{eq:Lsh}
\end{split}
\end{equation}
At early times $t \ll \tau$ the shock power rises as $L_{\rm sh} \propto t^{2}$, before peaking at $t_{\rm pk} \approx 1.1 \tau$ at a luminosity,
\be
\begin{split}
L_{\rm sh}^{\rm pk} & \approx \frac{1}{18}\frac{M_{\rm env} v_{\rm f}^{2}}{\tau} \\ & \approx 2.6\times 10^{38}\,{\rm erg\,s^{-1}}\,M_{\rm env, -4}\left(\frac{v_{\rm f, 8}}{2}\right) ^{2}\tau_{20}^{-1},
\label{eq:Lshmax}
\end{split}
\ee
and then decaying away exponentially away at late times $t \gg \tau$. Here, $M_{\rm env, -4} \equiv M_{\rm env}/(10^{-4}M_\odot)$, $v_{\rm f,8}\equiv v_{\rm f}/(1000 \rm \ km \ s^{-1})$, and $\tau_{20} \equiv \tau/(20\rm \ d)$.  

The strong shock heats gas to X-ray temperatures,
\be
T_{\rm sh}(t) \simeq \frac{3}{16}\frac{m_{\rm p}}{k}v_{\rm sh}^{2} \approx 2\times 10^{7}\,{\rm K} \,v_{\rm sh, 8}^{2},
\label{eq:Tsh}
\ee
compressing it to an immediate post-shock density
\be
\rho_{\rm sh} = 4\rho_{\rm w}(R_{\rm s}) \simeq \frac{\dot{M}_{\rm w}}{\pi R_{\rm s}^{2}v_{\rm w}},
\label{eq:rhosh}
\ee
where $\rho{\rm up} = \rho_{\rm w}(R_{\rm s}) = \dot{M}_{\rm w}/(4\pi R_{\rm s}^{2}v_{\rm w})$ is the upstream density of the shock and $v_{\rm sh, 8}\equiv v_{\rm sh}/(1000 \rm \ km \ s^{-1})$.

The shell is much cooler, with a temperature close to that achieved if the gas and radiation are in equilibrium, i.e.,
\be
\begin{split}
   T_{\rm s}(t) & \approx \left(\frac{L_{\rm sh} + L_{\rm wd}}{4\pi \sigma R_{\rm s}^{2}}\right)^{1/4} \\  \underset{\begin{subarray}{c} L_{\rm wd} \gg L_{\rm sh}, \\ t \gg \tau \end{subarray}} & \approx 1600\,{\rm K} \left(\frac{t}{\tau}\right)^{-1/2}\tau_{20}^{-1/2}\left(\frac{v_{\rm f, 8}}{2}\right) ^{-1/2}, 
   \label{eq:Ts}
\end{split}
\ee
where $L_{\rm wd} \approx L_{\rm Edd} \approx 1.4\times 10^{38}(M_{\rm wd}/M_{\odot})$ erg s$^{-1}$ is the intrinsic luminosity of the burning envelope, which we take equal the Eddington luminosity of the white dwarf of mass $M_{\rm wd} \approx 1M_{\odot}$ and in the final line we have used $R_{\rm s} \approx v_{\rm f}t/2$ for $t \gg \tau$ (Eq.~\eqref{eq:Rs}).  

The column density through the shell is given by
\be
\begin{split}
   N_{\rm H}(t) & = \frac{M_{\rm s}}{4\pi m_{\rm p} R_{\rm s}^{2}} \\  \underset{t \gg \tau}&\approx 3\times 10^{23}\,{\rm  cm^{-2}}\,M_{\rm env,-4}\left(\frac{v_{\rm f, 8}}{2}\right) ^{-2}\tau_{20}^{-2}\left(\frac{t}{\tau}\right)^{-2}, 
\end{split}
\label{eq:NH}
\ee
compatible with typical values $N_{\rm H} \sim 10^{22}-10^{24}$ cm$^{-2}$ inferred from X-ray spectra taken weeks to months into the outburst (e.g., \citealt{Orio+01,Mukai+14,Orio+15,Nelson+19}).

The mass density of the shell is related to its thickness $\Delta_{\rm s} \ll R_{\rm s}$ according to,
\be
\rho_{\rm s}(t) = \frac{M_{\rm s}}{4\pi \Delta_{\rm s} R_{\rm s}^{2}}.
\label{eq:rhos}
\ee
If the only source of pressure in the shell were the thermal pressure of the gas $P_{\rm g} \simeq \rho T_{\rm s}/\mu m_{\rm p}$, where $\mu \simeq 0.62$ is the mean molecular weight, then we would have a very thin-shell,
\be
\begin{split}
    \frac{\Delta_{\rm s}}{R_{\rm s}} \approx \mathcal{M}^{-2} \approx \frac{kT_{\rm s}}{\mu m_{\rm p} v_{\rm sh}^{2}} \approx 1.3\times 10^{-4}T_{\rm s,4}v_{\rm sh, 8}^{-2},
\end{split}
\ee
where $\mathcal{M} \approx v_{\rm sh}/c_{\rm s}$ is the Mach number of the shock and $T_{\rm s,4}\equiv T_{\rm s}/(10^4 \rm \ K)$.  In reality, several other effects likely prevent such extreme compression ratios, particularly the thin-shell instability \citep{Vishniac94,Steinberg&Metzger18}.  
The latter corrugates the shape of the shock front, which then injects vorticity into the downstream flow leading to turbulence. If a fraction $f_{\rm t} \ll 1$ of the shock luminosities $L_{\rm sh}$ feeds into turbulent pressure of the post-shock gas $P_{\rm t} \sim \rho_{\rm s}v_{\rm t}^{2}/2$, then
\be
\frac{\Delta_{\rm s}}{R_{\rm s}} \sim \frac{v_{\rm t}^{2}}{v_{\rm sh}^{2}} \sim 10^{-2}\left(\frac{f_{\rm t}}{10^{-2}}\right),
\label{eq:thickness}
\ee
where a value $f_{\rm t} \sim 10^{-2}$ is motivated by simulations of radiative shocks \citep{Steinberg&Metzger18} and X-ray observations fit to analytic mixing estimates (\citealt{Metzger+25}; see Eq.~\eqref{eq:fXmin} below).  A minimum shell thickness is also defined by the marginal stability criterion to the thin-shell instability \citep{Vishniac94,Steinberg&Metzger18},
which gives
\be
\frac{\Delta_{\rm s}}{R_{\rm s}} \sim \mathcal{M}^{-1} \sim 10^{-2}T_{\rm s,4}^{1/2}v_{\rm sh,8}^{-1}.
\label{eq:Deltas_thinshell}
\ee
Motivated thus, we hereafter assume a fiducial shell thickness $\Delta_{\rm s}/R_{\rm s} = 10^{-2}$, though most of our results are not sensitive to its precise value.  A thin shell geometry is consistent with optical spectral modeling of novae indicating a sharp outer ejecta density profile ($\rho \propto r^{-n}$ with $n \approx 15$; e.g., \citealt{Hauschildt+94a}).

\begin{figure}
    \centering
    \includegraphics[width=1.0\linewidth, clip=true,trim= 0 0 0 0]{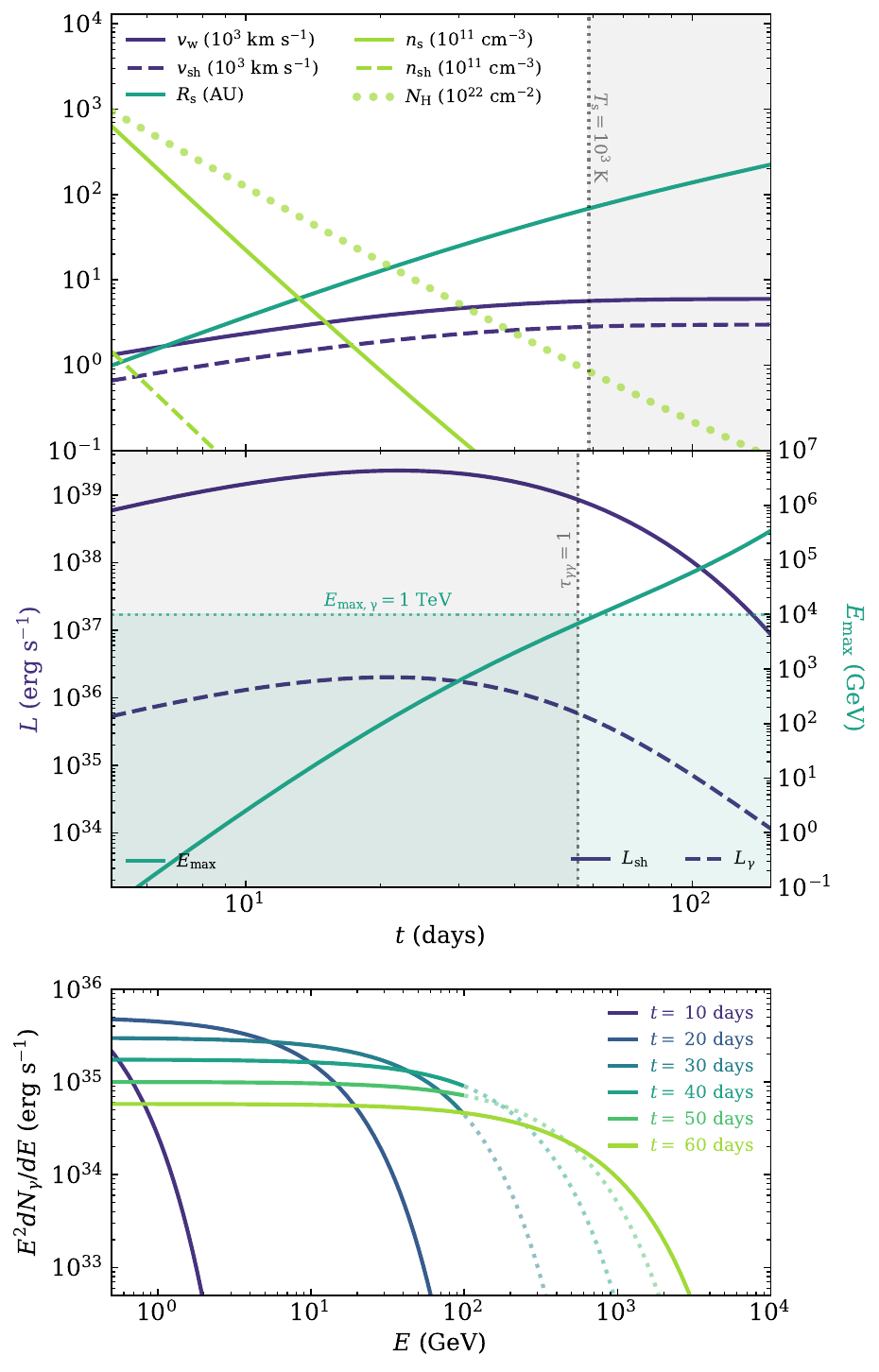}
    \caption{\emph{Top}: Hydrodynamic evolution of our fiducial nova with $M_{\rm env} = 10^{-4} M_{\odot}$, $v_{\rm f} = 6000 \rm{\ km \ s}^{-1}$, $\tau = 20$ days, $\Delta_{\rm sh}/R_{\rm s} = 10^{-2}$ (see Table \ref{tab:modelparams}). Note that the subscripts ``w", ``sh", and ``s" refer to the properties (velocity, radius, and number density) of the wind, shock, and shell, respectively; $N_{\rm H}$ refers to the column density through the shell. The vertical dotted line denotes the approximate time when the shell is cool enough to enable dust formation. \emph{Middle}: Evolution of the shock kinetic power ($L_{\rm sh}$), bolometric $\gamma$-ray luminosity ($L_\gamma$), and maximum proton energy ($E_{\rm max}$).  For the latter we assume $\xi_{\rm CR} = 0.03, \xi_{\rm B} = 0.01,f_{\Omega} = 0.3$. A vertical dotted line represents the time when the nova becomes optically thin to TeV $\gamma$-rays. Note that $E_{\rm max}$ continues to rise after the $\gamma$-ray peak, such that faint TeV $\gamma$-ray emission (corresponding to $E_{\rm max} \gtrsim 10$ TeV; horizontal dotted line) may be produced at late times. \emph{Bottom}: $\gamma$-ray spectra at selected epochs, assuming protons are accelerated with a distribution $\propto E^{-2}e^{-E/E_{\rm max}}$ (see Section \ref{subsec:spectra} details). Dotted lines indicate that the nova is optically thick to $\gtrsim 100$ GeV $\gamma$-rays (Sec.~\ref{sec:gammagamma}). }
    \label{fig:evolution}
\end{figure}

The resulting dense cool shell, so close to the shock, leads to efficient mixing between the hot and cold phases \citep{Metzger+25,Mitrani+25}, consistent with the clumpy nature of nova ejecta (e.g., \citealt{Williams13,Mason+18}) and the volume filling fraction for the line-emitting gas (e.g., \citealt{Ederoclite+06,Shore+13}).  The large reservoir of cool gas makes the shell an ideal site for dust nucleation \citep{Derdzinski+17}, and indeed many novae form dust starting soon after the most powerful GeV emission abates (\citealt{Chong+25}). Assuming most of the shock power $L_{\rm sh}$ is emitted as ionizing UV/X-ray radiation, the ionization parameter incident on the cold shell can be estimated:
\begin{eqnarray}
\xi_{\rm ion} & \equiv \frac{L_{\rm sh}}{n_{\rm s}R_{\rm s}^{2}} \underset{t \sim t_{\rm pk}} \approx \frac{2\pi}{9}\frac{M_{\rm env}}{M_{\rm s}(t_{\rm pk})}\frac{m_{\rm p}v_{\rm f}^{2}R_{\rm s}(t_{\rm pk})}{\tau}\left(\frac{\Delta_{\rm s}}{R_{\rm s}}\right) \nonumber
\\ & \approx 0.03\left(\frac{v_{\rm f, 8}}{2}\right) ^{3}\left(\frac{\Delta_{\rm s}}{10^{-2}R_{\rm s}}\right)\,{\rm erg\,cm\,s^{-1}}.
\end{eqnarray}
Here, $n_{\rm s} \equiv \rho_{\rm s}/m_{\rm p}$ and in the final equalities evaluated at peak shock power $t \approx t_{\rm pk}$ we have used $L_{\rm sh}^{\rm pk}$ (Eq.~\eqref{eq:Lshmax}), $R_{\rm s}(t_{\rm pk}) \approx 0.22v_{\rm f}\tau$, $M_{\rm s}(t_{\rm pk}) \approx 0.67 M_{\rm env}.$  The low value of $\xi_{\rm ion}$ implies that ionizing radiation is unlikely to penetrate the cool shell, which instead will remain sufficiently neutral to absorb and reprocess the shock's UV/X-ray emission (though photoionization may become important at late times as the shell density drops or in lower density polar regions of the wind; \citealt{Cunningham+15}).  Absent photo-ionization heating, the shell temperature will cool following Eq.~\eqref{eq:Ts}, enabling dust formation once the condensation temperature $T_{\rm s} \sim 10^{3}$ K is reached on a timescale $t\sim$ a few $\tau$ \citep[][see Figure \ref{fig:evolution} (top)]{Derdzinski+17}. 

Figure \ref{fig:evolution} (top) shows the hydrodynamic evolution for our fiducial model with $M_{\rm env} = 10^{-4} M_{\odot}$, $\tau = 20$ days, $v_{\rm f} = 6000 \rm{\ km \ s}^{-1}$, $\Delta_{\rm s}/R_{\rm s} = 10^{-2}$ (Table \ref{tab:modelparams}).  Both the wind and shock/shell velocities rise gradually to their maximum values over roughly a month, during which time the shock/shell radius accelerates.  The density of the shell is almost a thousand times larger than the immediate post-shock region at all times, but both decrease rapidly as the shell expands.  For the same model, Figure \ref{fig:evolution} (middle) shows the total shock power, which peaks around $t \sim \tau \sim 20$ days at $L_{\rm sh} \sim 3\times 10^{39}$ erg s$^{-1}$ (Eq.~\eqref{eq:Lshmax}).  The shock luminosity would be somewhat lower than this estimate if we had accounted for the smaller solid angle $f_{\Omega} < 1$ of the slow shell.  Nevertheless, the majority of the shock power is reprocessed into optical light \citep{Metzger+14}, whose peak luminosity can exceed the Eddington-limited luminosity of the white dwarf (shedding light on the long-standing mystery of super-Eddington novae; e.g., \citealt{Shaviv01,Kato&Hachisu05,Kato&Hachisu07}).

\subsection{$\gamma$-ray luminosity}

We consider that cosmic rays are accelerated at the shock with an efficiency $\xi_{\rm CR} < 1$ and a luminosity $L_{\rm CR} = \xi_{\rm CR}L_{\rm sh}$, after which they are effectively immediately advected downstream into the shell. Assuming the magnetic field in the shell is sufficient to confine the cosmic rays, their total energy, $E_{\rm CR}$, will evolve in time as:
\be
\frac{dE_{\rm CR}}{dt} = L_{\rm CR} - \frac{E_{\rm CR}}{t_{\pi}} - \frac{E_{\rm CR}}{t},
\label{eq:dECRdt}
\ee
where $t_{\pi} = (n_{\rm s}\sigma_{\rm pp}c)^{-1}$ is the pion-creation timescale, where $\sigma_{\rm pp} \approx 5\times 10^{-26}$ cm$^{2}$ and the final term $P_{\rm CR}(dV_{\rm s}/dt) = E_{\rm CR}/t$ accounts to PdV adiabatic losses, where $P_{\rm CR} = E_{\rm CR}/(3 V_{\rm s})$ is the cosmic ray pressure (treated as a fluid with an adiabatic index $\gamma_{\rm ad} = 4/3$) and $V_{\rm s} \propto R_{\rm s}^{3}$ is the shell volume, assuming three-dimensional expansion (i.e., that the shell thickness scales with the shell radius, $\Delta_{\rm s} \propto R_{\rm s}$; the prefactor would change slightly if $\Delta_{\rm s} = const$ such that the shell expansion was effectively 2D).\footnote{Equation \eqref{eq:dECRdt} neglects any boost to the CR energy as the result of their compression with the thermal plasma as the latter is incorporated into the cool shell, which in principle could boost $E_{\rm CR}$ by a factor of $\sim (\rho_{\rm s}/\rho_{\rm sh})^{1/3} \sim 10$ treating the CR as a $\gamma = 4/3$ gas \citep{Vurm&Metzger18}; this effect could be roughly incorporated into the model as an effective increase in the shock acceleration efficiency $\xi_{\rm CR}$.}

Given a solution to Eq.~\eqref{eq:dECRdt}, the emitted $\gamma$-ray luminosity can be written 
\be
L_{\gamma} = \frac{f_\Omega\kappa E_{\rm CR}}{t_{\pi}},
\ee
where $\kappa \simeq 0.1$ is the fraction of the cosmic ray energy that is radiated as $\gamma$-rays through p-p interactions and $f_{\Omega} \equiv \Delta \Omega/4\pi \gtrsim 0.1$ is the covering fraction of the radiative shock, assuming the slow shell subtends a solid angle $\Delta \Omega$ (Fig.~\ref{fig:diagram}).

There are two limits, depending on the ratio of the two energy loss terms in Eq.~\eqref{eq:dECRdt},
\be
\frac{t_{\pi}}{t} = \frac{1}{n_{\rm s}\sigma_{\rm pp}c t} = \frac{4\pi m_{\rm p}}{M_{\rm s} \sigma_{\rm pp}}\frac{R_{\rm s}^{3}}{c t}\frac{\Delta_{\rm s}}{R_{\rm s}}
\label{eq:tratio}
\ee
When $t_{\pi} \ll t$ (``calorimetric'' limit), then the steady-state solution to Eq.~\eqref{eq:dECRdt} for $dE_{\rm CR}/dt \approx 0$ is given by $E_{\rm CR} \approx L_{\rm CR}t_{\pi}$, and hence
\be
L_{\gamma} \approx f_\Omega \xi_{\rm CR} \kappa L_{\rm sh},
\label{eq:Lgamma}
\ee
i.e., the $\gamma$-ray luminosity faithfully tracks the instantaneous shock power, and hence the portion of the UV/optical emission due to reprocessed shock emission (e.g., \citealt{Metzger+15}).  This is consistent with the roughly one-to-one mapping observed between the observed duration of the $\gamma$-ray emission, $t_{\gamma}$, in classical novae and the speed class $t_{2}$ \citep{Franckowiak+18,Craig+25}, because in our toy model $L_{\rm sh}$ and hence $L_{\gamma}$ peaks over a characteristic duration $t_{\gamma} \sim 2t_{\rm pk} \sim 2\tau \sim t_{2}$ (Fig.~\ref{fig:evolution}) provided that the nova optical light curve decline indeed tracks the mass-loss rate of the white dwarf (Eq.~\eqref{eq:Mdotw}).  In novae for which temporally correlated optical/$\gamma$-ray emission is observed, the measured ratio $L_{\rm opt}/L_{\gamma}$ constrains $\xi_{\rm CR} \kappa \sim 10^{-2}-10^{-3}$ \citep{Metzger+15,Li+17,Aydi+20} and thus $\xi_{\rm CR} \sim 0.01-0.1$ for $\kappa \approx 0.1.$  A modest proton acceleration efficiency $\sim 1\%$ is consistent with that expected because the magnetic field of the unshocked nova wind is predominantly toroidal and hence perpendicular to the radial shock normal \citep{Steinberg&Metzger18,Orusa+25b}. 

Considering Eq.~\eqref{eq:tratio} at the time of peak shock power ($t_{\rm pk} \approx 1.1\tau$; Eq.~\eqref{eq:Lshmax}) we have
\be
\begin{split}
\left.\frac{t_{\pi}}{t}\right|_{t_{\rm pk}} &= \frac{4\pi m_{\rm p}}{M_{\rm s}(t_{\rm pk}) \sigma_{\rm pp}}\frac{R_{\rm s}^{3}(t_{\rm pk})}{c t_{\rm pk}}\frac{\Delta_{\rm s}}{R_{\rm s}} \\
&\approx 2.4\times 10^{-4}M_{\rm env, -4}^{-1}\left(\frac{\Delta_{\rm s}/R_{\rm s}}{10^{-2}}\right)\left(\frac{v_{\rm f, 8}}{2}\right) ^{3}\tau_{20}^{2},
\end{split}
\label{eq:tpiratio}
\ee
where we have again used the fact that $M_{\rm s}(t_{\rm pk}) \approx 0.67 M_{\rm s}$ and $R_{\rm s}(t_{\rm pk}) \approx 0.22 v_{\rm f}\tau$.  Thus, for typical ejecta masses and ejecta velocities, we are safely in the calorimetric limit at peak shock power, as long as the shell is thin as expected $\Delta_{\rm s}/R_{\rm s} \sim 10^{-2}$ (Eq.~\eqref{eq:Deltas_thinshell}).  The maximum $\gamma$-ray luminosity is then
\begin{eqnarray}
&&L_{\gamma}^{\rm pk} \approx f_\Omega\xi_{\rm CR} \kappa L_{\rm sh}^{\rm pk}  \nonumber \\
& \approx& 2.6\times 10^{35} {\rm erg\,s^{-1}\,}\left(\frac{f_\Omega\xi_{\rm CR} \kappa}{10^{-3}}\right)\frac{M_{\rm env, -4}}{\tau_{\rm 20}}\left(\frac{v_{\rm f, 8}}{2}\right) ^{2}, \nonumber \\
\label{eq:Lgammamax}
\end{eqnarray}
where we have used Eq.~\eqref{eq:Lshmax}.  

At late times $t \gg \tau$ the ratio $t_{\pi}/t \propto t^{2}$ increases until the cosmic rays enter the ``adiabatic'' limit ($t_{\pi} \gtrsim t$), after a time: 
\be
\begin{split}
\frac{t_{\rm ad}}{\tau} & \equiv \left(\frac{2}{\pi}\frac{M_{\rm s}\sigma_{\rm pp}}{m_{\rm p}}\frac{c}{v_{\rm f}^{3}\tau^{2}}\frac{R_{\rm s}}{\Delta_{\rm s}}\right)^{1/2} \\ & \approx 22 \ M_{\rm env, -4}^{1/2}\left(\frac{v_{\rm f, 8}}{2}\right) ^{-3/2}\tau_{20}^{-1}\left(\frac{\Delta_{\rm s}}{10^{-2}R_{\rm s}}\right)^{-1/2},
\end{split}
\ee
where we take $R_{\rm s} \approx v_{\rm f}t/2$ and $M_{\rm s} \simeq M_{\rm env}$ at times $t \approx t_{\rm ad} \gg \tau$.  At times $t \gtrsim t_{\rm ad}$, adiabatic losses become important and the $\gamma$-ray luminosity no longer tracks the shock power.  

The predicted bolometric $\gamma$-ray luminosity of our fiducial nova model is shown as a dashed curve in Figure \ref{fig:evolution} (middle), which we have obtained by solving Eq.~\eqref{eq:dECRdt} assuming $f_{\Omega} = 0.3, \xi_{\rm CR} = 0.03, \kappa = 0.1$.


\subsection{Maximum Particle Energy}
\label{subsec:emax}

We assume that particle acceleration is limited to the hot ionized region around the shock. In the upstream, UV and X-ray photons from the shock itself serve as an ionization source and set the extent of the acceleration region \citep{Metzger+16}. In the downstream, the acceleration region extends from the shock to the cool, partially neutral shell. In other words, particles that diffuse far enough to reach the dense shell cannot return to the shock and, in effect, ``escape." This is motivated by the short pion-creation cooling timescale that relativistic ions experience once they enter the dense shell (see Eq.~\eqref{eq:tpiratio}). 

If the post-shock flow were laminar, then the thickness of the immediate downstream region, $\Delta_{\rm down}$, is set by the length, $\Delta_{\rm rad}$, over which the hot shocked gas of temperature $T_{\rm sh} \gtrsim 10^{7}$ K (Eq.~\eqref{eq:Tsh}) cools radiatively:
\be
\begin{split}
\frac{\Delta_{\rm down}}{R_{\rm s}} & \sim \frac{\Delta_{\rm rad}}{R_{\rm s}} \approx \frac{v_{\rm down}t_{\rm cool}}{R_{\rm s}} \\ & \approx \frac{3}{16}\frac{kT_{\rm sh} v_{\rm w}}{n_{\rm sh}\Lambda(T_{\rm sh})R_{\rm s}} = \frac{3\pi}{16}\frac{kT_{\rm sh}m_{\rm p} v_{\rm w}^{2}R_{\rm s}}{\dot{M}_{\rm w}\Lambda(T_{\rm sh})} \\
 \underset{t \gg \tau} & \approx 5.29\times10^{-2}
M_{\rm env, -4}^{-1}
\tau_{20}^{2}
v_{\rm sh,8}^{4}
\left(\frac{t}{\tau}\right)e^{t/\tau},
\end{split}
\label{eq:Deltarad}
\ee
where $v_{\rm down} \approx v_{\rm sh}/4 = v_{\rm w}/8$ and $n_{\rm sh} = \rho_{\rm sh}/m_{\rm p}$ (Eq.~\eqref{eq:rhosh}) are the immediate post-shock downstream velocity and density and $\Lambda(T_{\rm sh}) \approx \Lambda_0T_{\rm sh}^{1/2}\approx2\times 10^{-27}T_{\rm sh}^{1/2}$ erg cm$^{3}$ s$^{-1}$ is the free-free cooling function.  

This relatively short cooling length, $\Delta_{\rm rad} \lesssim R_{\rm s}$, was used to argue that nova reverse shocks are often at least marginally radiative \citep{Metzger+14}.  However, if direct radiative cooling of the hot gas were indeed efficient, then a significant fraction of the shock power $L_{\rm sh} \gtrsim 10^{37}-10^{38}$ erg s$^{-1}$ (Eq.~\eqref{eq:Lshmax}) should be emitted as thermal X-rays of temperature $kT_{\rm sh} \sim 1-10$ keV (Eq.~\eqref{eq:Tsh}).  This conflicts with X-ray observations of classical novae which show $f_{\rm X} \equiv L_{\rm X}/L_{\rm sh} \sim 10^{-5}-10^{-4}$ \citep{Nelson+19,Sokolovsky+20,Sokolovsky+22}, thus revealing the true thickness of the X-ray emitting layer, $\Delta_{\rm down}$, to be much smaller than $\Delta_{\rm rad}$.  In particular, following \citet[][their Eq.~12]{Metzger+25}, one can translate an empirically observed X-ray efficiency into the hot layer thickness according to:
\be
\Delta_{\rm down} \approx (32/9)f_{\rm X}f_{\Omega}^{-1}\Delta_{\rm rad} \ll \Delta_{\rm rad}.
\label{eq:deltaDS}
\ee
Here, the factor $f_{\Omega}^{-1} > 1$ accounts for the fact that if the shocks cover only a fraction of the solid angle, their radial thickness must be greater to generate a given observed X-ray luminosity than in the spherical shock case.  \citet{Metzger+25} argue that the suppression $f_{\rm X} \ll 1$ results from mixing of the hot X-ray emitting gas with the cool shell, due to turbulence (e.g., driven by the thin-shell instability; \citealt{Steinberg&Metzger18}).\footnote{Assuming Komolgorov scalings for the post-shock turbulence, \citet{Metzger+25} derive a {\it minimum} value for $f_{\rm X}$ (their Eq.~23) given by
\be
f_{\rm X,min} \approx 5\times 10^{-5}\left(\frac{f_{\rm t}}{10^{-2}}\right)^{-3/2}v_{\rm sh,8}^{-3},
\label{eq:fXmin}
\ee
where again $f_{\rm t}$ is the fraction of the shock power placed into turbulence (e.g., \citealt{Steinberg&Metzger18}).} 

In what follows, we use $\Delta_{\rm down}$ from Eq.~\eqref{eq:deltaDS} for an assumed constant value of $f_{\rm X} \sim 10^{-4}$ motivated by nova observations (see Sec.~\ref{sec:fX} for a discussion).  Under this assumption, and using Eq.~\eqref{eq:Deltarad}, we see that the width of the post-shock region $\Delta_{\rm down}/R_{\rm s} \propto v_{\rm w}^{3} \propto v_{\rm sh}^{3}$ grows rapidly as the wind and shock velocity rises.  This has important implications for the predicted evolution of the $\gamma$-ray emission, as we now discuss.



Assuming particle acceleration is limited to the postshock region of thickness $\Delta_{\rm down}$, we estimate the maximum proton energy, $E_{\rm max}$ by requiring that the downstream diffusion length $\ell_{\rm diff,ds}$ be smaller than $\Delta_{\rm down}$. In other words, we have,
\begin{equation}
    \ell_{\rm diff,ds} = \frac{D_{\rm down}}{v_{\rm down}} \approx \frac{4cr_{\rm L}(E_{\rm max})}{3v_{\rm sh}} = \Delta_{\rm down}.
    \label{eq:Emaxcondition}
\end{equation}
Here, we assume Bohm diffusion and $r_{\rm L}(E_{\rm max})$ is the gyroradius of protons with energy $E = E_{\rm max}$, given by 
\begin{equation}
    r_{\rm L}(E_{\rm max}) \approx 3.3\times10^6 \text{ cm }\bigg(\frac{E_{\rm max}}{\rm GeV}\bigg)\bigg(\frac{B}{\rm G}\bigg)^{-1}.
    \label{eq:rL}
\end{equation}

Even a strong magnetic field on the white dwarf surface of radius $R_{\rm WD} \lesssim 10^{9}$ cm is diluted by flux-freezing to negligibly small values by the radius of the shock, $R_{\rm s} \gtrsim 10^{13}-10^{14}$ cm (e.g., \citealt{Metzger+15}).  This suggests the magnetic fields responsible for particle acceleration are generated locally at the shock (e.g., \citealt{Li+17}; Sec.~\ref{sec:Bfield}). If a fraction $\xi_{\rm B}$ of the shock's ram pressure ($\approx\rho_{\rm sh}v_{\rm sh}^2/4$) is placed into the energy of the magnetic field, this results in a magnetic field strength:
\be
B_{\rm down} \approx \left(\xi_{\rm B}\frac{\dot{M}_{\rm w}v_{\rm sh}}{R_{\rm s}^{2}}\right)^{1/2}.
\ee
From Eq.~\eqref{eq:Emaxcondition}, this results in a maximum proton energy:
\be
\begin{split}
E_{\rm max} & = \frac{3}{4}\frac{v_{\rm sh}}{c}eB_{\rm down}\Delta_{\rm down} \\ & \approx 2.7 \frac{e}{c}\frac{k^{1/2}m_{\rm p}^{3/2}}{\Lambda_0}\xi_{\rm B}^{1/2}f_{\rm X}f_\Omega^{-1}\frac{v_{\rm sh}^{9/2}R_{\rm s}}{\dot{M}_{\rm w}^{1/2}} \\ 
 \underset{t \sim t_{\rm pk}}&{\approx} 12\,{\rm GeV}\,\\ &\times f_{\rm X,-4}f_{\Omega,-1}^{-1}\xi_{\rm B,-2}^{1/2}v_{\rm sh, 8}^{11/2}M_{\rm env,-4}^{-1/2}\tau_{20}^{3/2}.
\end{split}
\label{eq:Emax}
\ee
Here, $f_{\rm X,-4} \equiv f_{\rm X}/10^{-4}$, $f_{\Omega,-1} \equiv f_{\Omega}/0.1$, $\xi_{\rm B,-2} \equiv \xi_{\rm B}/10^{-2}$, and in the final line we have used $R_{\rm s}(t_{\rm pk}) \approx 0.2 v_{\rm f}\tau \approx 0.4v_{\rm sh}\tau$ and $\dot{M}_{\rm w}(t_{\rm pk}) \approx 0.33M_{\rm env}/\tau$.

Because $E_{\rm max} \sim 10E_{\gamma}$ is required to generate a $\gamma$-ray of energy $\sim E_{\gamma}$ via pion decay, we see that emission in the \emph{Fermi} LAT band, $E_{\gamma} \gtrsim $GeV, is possible near maximum shock power for typical nova outflow properties.  At times $t \gg \tau,$ $\dot{M}_{\rm w}$ drops as $\propto e^{-t/\tau}$, while $v_{\rm sh}$ increases, and hence $E_{\rm max}$ grows rapidly in time (Sec.~\ref{sec:TeV}).

Expressed in terms of the $\gamma$-ray luminosity, $L_{\gamma} = (3/2)\xi_{\rm CR}\kappa \dot{M}_{\rm w}v_{\rm sh}^{2}$, one finds
\be
\begin{split}
E_{\rm max} &\approx 3.3 \frac{e}{c} f_{\rm X}f_{\Omega}^{-1} \frac{k^{1/2}m_{\rm p}^{3/2}\xi_{\rm B}^{1/2}\xi_{\rm CR}^{1/2}\kappa^{1/2}}{\Lambda_0}\frac{v_{\rm sh}^{11/2}R_{\rm s}}{L_{\gamma}^{1/2}} \\
\underset{t \sim t_{\rm pk}}&{\approx} 30\,{\rm GeV}\, \\ &\times f_{\rm X,-4}f_{\Omega,-1}^{-1/2}\xi_{\rm B,-2}^{1/2}\xi_{\rm CR,-2}^{1/2}v_{\rm sh,8}^{13/2}\tau_{20}L_{\gamma,35}^{-1/2},
\label{eq:Emax_Lgamma}
\end{split}
\ee
where $\xi_{\rm CR,-2} \equiv \xi_{\rm CR}/0.01$ and $L_{\gamma,35} \equiv L_{\gamma}/(10^{35} \rm{\ erg \ s^{-1}})$.   The final line estimates $E_{\rm max}$ at the peak of the shock power ($t \sim t_{\rm pk})$; however, the extremely strong dependence on $v_{\rm sh}$ suggests that, after $t_{\rm pk}$, $E_{\rm max}$ continues to grow rapidly.  

The time evolution of $E_{\rm max}$ for our fiducial model is shown in Figure \ref{fig:evolution} (bottom), for $\xi_{\rm B} = 0.01$, $f_{\Omega} = 0.3$, $f_{\rm X} = 5\times10^{-5}$.  We see that $E_{\rm max}$ rises from $\sim 100$ GeV at $t_{\rm pk} \sim \tau \sim 20$ d to $E_{\rm max} \gtrsim 10$ TeV over just a couple months. The observational implications of this predicted evolution will be discussed in Section \ref{sec:results}.

\subsection{$\gamma$-ray Spectra} 
\label{subsec:spectra}
The formalism described in the preceding sections naturally lends itself to multi-zone predictions of the $\gamma$-ray spectrum, as shown in Figure \ref{fig:evolution} (bottom) for our fiducial nova. To estimate $dN_\gamma/dE$, we assume that during each epoch, $t_{\rm i}$, spanning an interval $\delta t_{\rm i}$, the shock instantaneously accelerates a power-law proton spectrum $\phi_{\rm i}(E, t_{\rm i}) = A_{\rm i}(E/E_0)^{-2}e^{-E/E_{\rm max}(t_{\rm i})}$, where $E_0$ is an arbitrary energy scaling and the normalization, $A_{\rm i}$, is set such that,
\begin{equation}
    \int_{m_p c^2}^{\infty}E\phi_{\rm i}(E, t_{\rm i})dE = L_{\rm CR}(t_{\rm i})\delta t_{\rm i}.
\end{equation}
From the time of acceleration ($t_{\rm i}$), to the current time, $t_{\rm now}$, this instantaneous spectrum is subject to adiabatic losses approximated by $L_{\rm ad}(t_{\rm i},t_{\rm now})\equiv R_{\rm s}(t_{\rm now})/R_{\rm s}(t_{\rm i})\geq 1$ and proton-proton losses aproximated by $L_{\rm \pi}(t_{\rm i},t_{\rm now}) \equiv \exp{[-(t_{\rm now}-t_{\rm i})/t_\pi^{\rm ave}]}\leq 1$. Here, $t_\pi^{\rm ave}$ is the average $t_\pi$ from $t_{\rm i}$ to $t_{\rm now}$. The evolved instantaneous proton spectrum thus becomes,
\begin{equation}
   \phi_{\rm i}(E,t_{\rm now}) = L_{\rm ad}(t_{\rm i},t_{\rm now})L_\pi(t_{\rm i},t_{\rm now}) \phi_{\rm i}(E',t_{\rm i}),
\end{equation}
where $E'\equiv L_{\rm ad}(t_{\rm i},t_{\rm now})E$ accounts for the shift in energy due to adiabatic expansion.

Thus, the cumulative proton spectrum is simply $\phi(E) = \sum_{i}\phi_{\rm i}(E, t_{\rm now})$, and the corresponding differential $\gamma$-ray flux is
\be
\frac{dN_{\rm \gamma}}{dE} = f_{\Omega}\frac{\phi(E/\kappa)}{\kappa t_\pi^{\rm now}}.
\ee
Note that integrating this spectrum gives a bolometric $L_{\gamma}$ that is within 1\% of that calculated by solving Equation \eqref{eq:dECRdt}.  Snapshots of the spectra calculated for our fiducial model are shown in the bottom panel of Fig.~\ref{fig:evolution}.

\begin{figure*}
    \centering
    \includegraphics[width=0.99\linewidth]{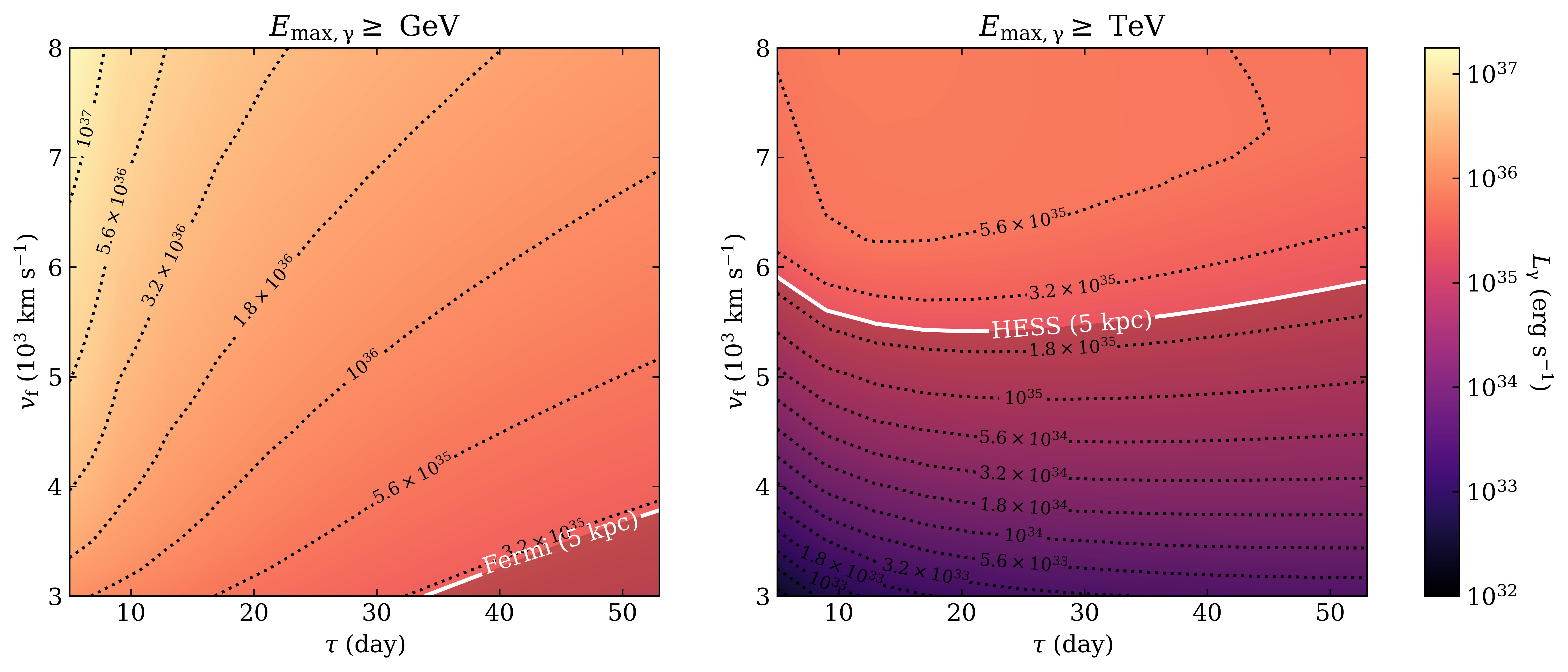}
    \caption{Predicted maximum $\gamma$-ray luminosity (color scale) as a function of final wind velocity ($v_{\rm f}$) and characteristic timescale ($\tau$), assuming the maximum $\gamma$-ray energy, $E_{\rm max,\gamma}$, exceeds 1 GeV (left panel) or 1 TeV (right panel). For the right (TeV) panel, we also require that the nova be optically thin ($\tau_{\gamma\gamma} < 1$). Note that $\tau$ and $v_{\rm f}$ are related to the speed class, $t_2$, and the maximum inferred ejecta velocity, $v_2$, as $\tau\approx 0.5t_2$ and $v_{\rm f} \approx 2v_2$. We assume a fixed ejecta mass $M_{\rm env} = 10^{-4}M_{\odot}$ and all other parameters are set to their fiducial values (Table~\ref{tab:modelparams}). Approximate \emph{Fermi} LAT and H.E.S.S. $5\sigma$ detection thresholds for sources at distance $d = 5$ kpc are overlaid (white lines).}
    \label{fig:luminosities}
\end{figure*}

\subsection{$\gamma$-ray Absorption}
\label{sec:gammagamma}

$\gamma$-rays produced at the shock can in principle be attenuated as they escape to the distant observer due to interaction with matter or radiation ahead of the shock.  $\gamma$-rays interact with the nuclei in the ejecta (of atomic mass $A$ and charge $Z$) producing electron/positron pairs through the Bethe–Heitler (BH) process.  The cross section for this process can be approximated as \citep{Chodorowski+92}
\be
\sigma_{\rm BH} \simeq \frac{3}{8\pi}\alpha \sigma_{\rm T}Z^{2}\left[\frac{28}{9}\ln\left(\frac{2E_{\gamma}}{m_{\rm e} c^{2}}\right) - \frac{218}{27}\right],
\ee
where $\alpha \simeq 1/137$ and $\sigma_{\rm T} = 6.65\times 10^{-25}$ cm$^{2}$.  This gives a BH optical depth through the cool shell:
\be
\begin{split}
\tau_{\rm BH} = \frac{\sigma_{\rm BH}N_{\rm H}}{A}  \approx 0.045\frac{Z^{2}}{A} \left(\frac{\ln\left[4\times 10^{6}(E_{\rm \gamma, GeV})\right]}{15.2}\right) \\ \times \  
  M_{\rm s,-4}\left(\frac{R_{\rm s}}{0.22 v_{\rm f}\tau}\right)^{2}\left(\frac{v_{\rm f,8}}{2}\right)^{-2}\tau_{20}^{-2},
\end{split}
\ee
where $E_{\gamma, \rm GeV} \equiv E_{\gamma}/\rm{GeV}$, $M_{\rm s,-4}\equiv M_{\rm s}/10^{-4}M_\odot$, and we have estimated the particle column using Eq.~\eqref{eq:NH}, normalizing the shell radius to its value at peak luminosity.  Even assuming the ejecta composition to be dominated by CNO nuclei ($A = 2Z = 12-16$), the shell is unlikely to be opaque to $\sim$ GeV photons for $t\gtrsim t_{\rm pk}$.

Higher-energy $\gamma$-rays ($E_{\gamma} \sim$ TeV) can also interact with ambient photons to create electron/positron pairs. Assuming that most of the shock power is ultimately reprocessed into optical emission, the optical radiation energy density from this emission can be written,
\be
u_{\gamma} \simeq \frac{L_{\rm sh}}{4\pi c R_{\rm s}^{2}}.
\ee
The optical depth for a TeV photon to interact with the nova optical light is thus given by
\begin{equation}
\begin{split}
\tau_{\gamma\gamma} &\simeq \frac{u_{\gamma}}{E_{\rm opt}}R_{\rm s}\sigma_{\gamma\gamma} \underset{t \gg \tau}\approx \frac{3}{64\pi}\frac{M_{\rm env}v_{\rm f}}{c\tau^{2}(t/\tau-1)}\frac{\sigma_{\rm T}}{E_{\rm opt}}e^{-t/\tau} \\
 \underset{t \gg \tau}&{\approx}  11 M_{\rm env, -4}\left(\frac{v_{\rm f, 8}}{2}\right) \tau_{20}^{-2}\frac{e^{-t/\tau}}{(t/\tau-1)},
\end{split}
\label{eq:taugg}
\end{equation}
where $E_{\rm opt} \approx m_{e}^{2}c^{4}/E_{\gamma} \approx 0.25$ eV is the energy of the IR/optical seed photons that will pair-create on photons of energy $E_{\gamma} \approx $ TeV and $\sigma_{\gamma\gamma} \approx \sigma_{\rm T}/4$ \citep{Berestetskii+82}, where $\sigma_{\rm T}$ is the Thomson cross section. 

Thus, $\gamma\gamma$ absorption can be important at attenuating TeV photons near the peak of the shock power.  However, unless $t\gg\tau$, $\tau_{\gamma\gamma}$ is somewhat lower than the estimate given in the final line of Eq.~\eqref{eq:taugg}. Our fiducial nova becomes optically thin to high-energy $\gamma$-rays ($\tau_{\gamma\gamma}\lesssim 1$) at approximately $2.5\tau$, approximately when $E_{\rm max}$ reaches 10 TeV (vertical dotted line in the bottom panel of Fig.~\ref{fig:evolution}). 

Note that the energy dependence of $\tau_{\gamma\gamma}$ depends on the distribution of seed photons. However, under our assumption that most of the shock emission is optical, we do not expect $\tau_{\rm \gamma\gamma}$ to be significant for $\gamma$-rays with energies $\lesssim 100$ GeV.

\section{Implications for $\gamma$-ray Observations} 
\label{sec:results}

In this Section, we discuss implication of our model for $\gamma$-ray observations of novae. Throughout, we will approximate the \emph{Fermi} LAT detection threshold as $F_{\gamma} \equiv L_\gamma/(4\pi d^2)\gtrsim 10^{-10} \rm{ \ erg \ cm^{-2} \ s^{-1}}$, where $d$ is the source distance. While the precise \emph{Fermi} LAT sensitivity depends on integration time and Galactic latitude \citep{Atwood+09}, this value is roughly consistent with the faintest $\gamma$-ray detected novae \citep{Craig+25}.

As we will demonstrate, a subset of \emph{Fermi}-detectable novae may also be detectable at TeV energies with current IACTs. Using H.E.S.S. as a fiducial example of the latter, we take the one-hour detection threshold to be roughly 0.05 times the flux of the Crab pulsar above 1 TeV \citep{Aharonian+08}. Assuming a $\gamma$-ray spectrum of the form $f(E) \propto E^{-2}e^{-E/E_{\rm max,\gamma}}$, and taking $E_{\rm max, \gamma} = 1$ TeV, we thus require $F_\gamma \gtrsim 7.5\times10^{-11} \rm \ erg \ cm^{-2} \ s^{-1}$. 

\begin{figure*}
    \centering
    \includegraphics[width=0.5\linewidth, clip=true,trim= 0 0 0 0]{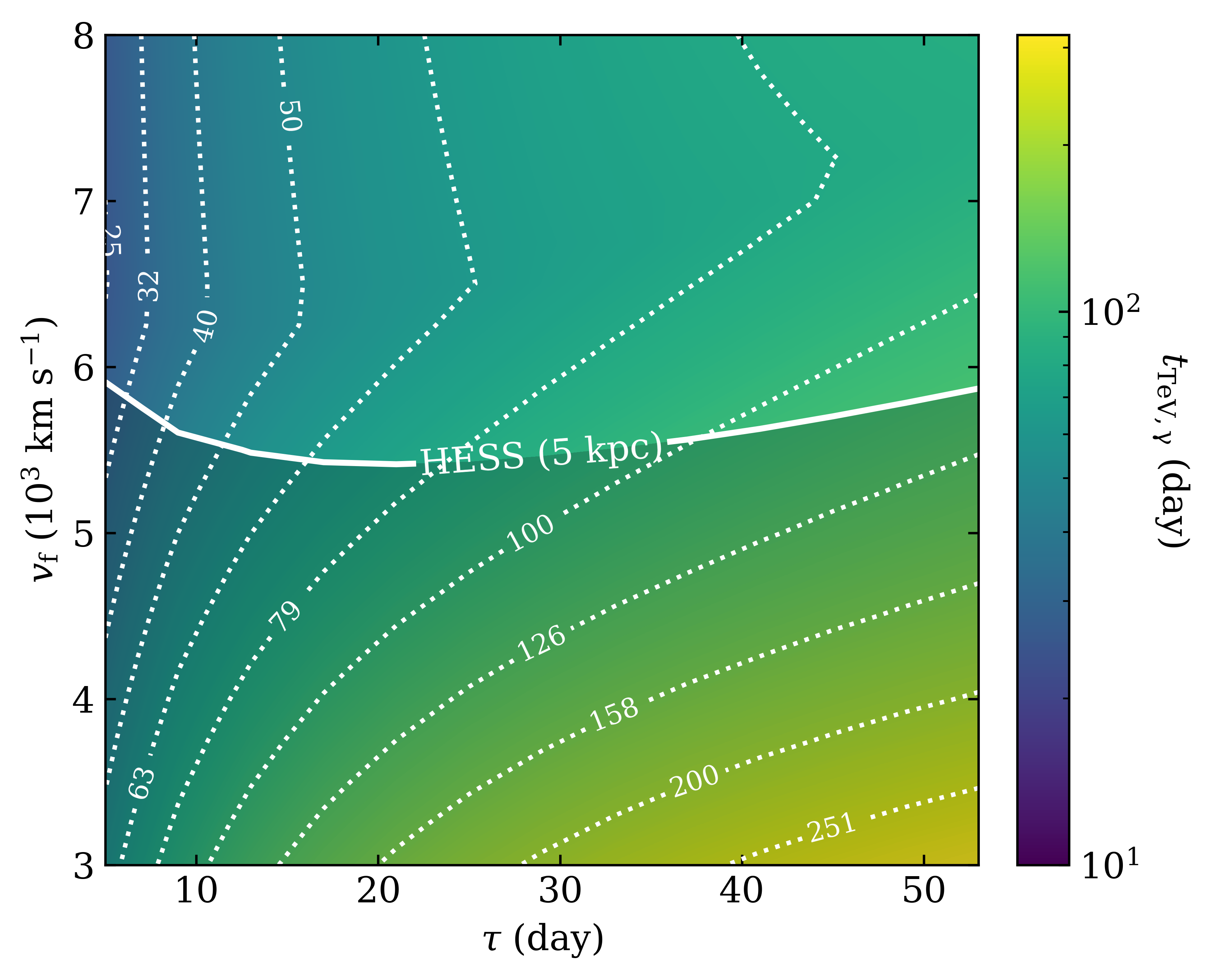}
    \includegraphics[width=0.49\linewidth, clip=true,trim= 0 0 0 0]{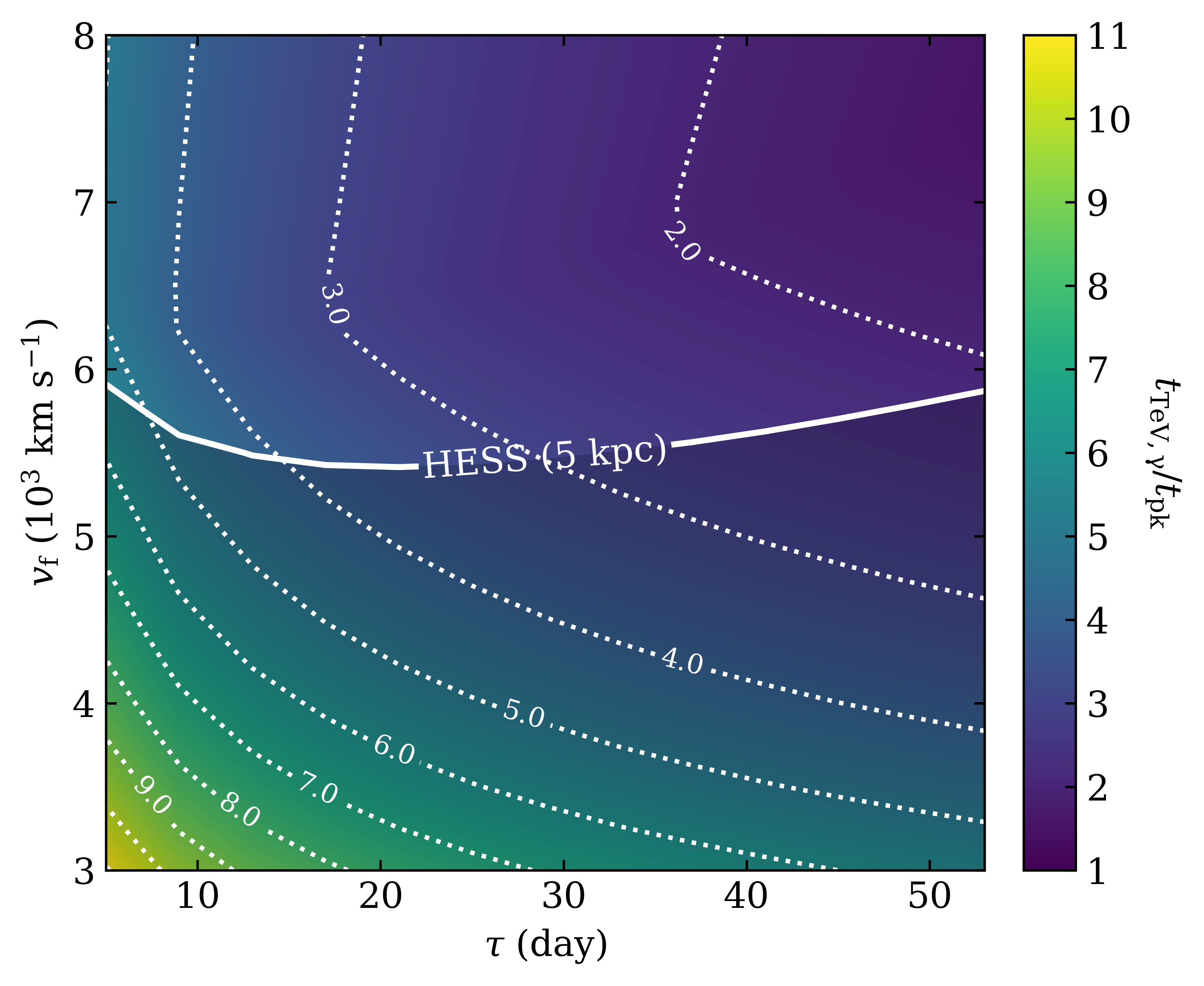}
    \caption{\emph{Left:} Time after eruption in delays when a nova is first expected to produce unabsorbed TeV $\gamma$-rays ($t_{\rm TeV,\gamma}$, color scale) as a function of final wind velocity ($v_{\rm f}$) and characteristic timescale ($\tau$). All other parameters are set according to our fiducial nova. The approximate H.E.S.S. detection threshold (see Figure \ref{fig:luminosities}) is overlaid. \emph{Right:} Same as left, but expressed as a multiple of $t_{\rm pk}$. Note that some novae may be detectable at TeV energies but, taking $\tau\approx 0.5t_2$, only after days or even weeks past the peak of the optical/$\gamma$-ray emission.}
    \label{fig:ttev}
\end{figure*}

\subsection{Peak GeV Emission}
\label{sec:GeV}

In the parameter space of $\tau$ (a proxy for nova speed class) and final wind speed $v_{\rm f} \approx 2v_{2}$, Figure \ref{fig:luminosities} shows contours of maximum $\gamma$-ray luminosity at times when $E_{\rm max,\gamma} = E_{\rm max}/10$ exceeds 1 GeV (left panel) and 1 TeV (right panel), respectively.  Here, we have fixed $M_{\rm env} = 10^{-4}M_{\odot}$ and other parameters at their fiducial values (Table \ref{tab:modelparams}).  In the right panel (and in subsequent figures), we account for absorption of TeV $\gamma$-rays by only considering times when $\tau_{\gamma\gamma}<1$.

We see that a subset of nearby ($d\lesssim 5$ kpc) classical novae are detectable at GeV energies with \emph{Fermi} LAT: those with short characteristic timescales (i.e., small $t_2$) and/or high outflow velocities. The parameters of our theoretically detectable novae are broadly consistent with the population that \emph{Fermi} LAT has actually observed, most of which have $v_{2} \approx v_{\rm f}/2 \gtrsim 2000$ km s$^{-1}$ and $t_{2} < 50$ day (\citealt{Craig+25}).  However, as Figure \ref{fig:luminosities} assumes a fixed $M_{\rm env}$, and since $L_{\gamma} \propto M_{\rm env}$ (Eq.~\eqref{eq:Lgammamax}), the precise subset of detectable $\tau$ and $v_{\rm f}$ can vary. 

Moreover, if the GeV emission is sufficiently luminous to be detected ($L_{\gamma} \gtrsim 3\times 10^{35}$($d$/5 kpc)$^{2}$ erg s$^{-1}$) then the timescale for the emission to reach GeV energies (i.e., $E_{\rm max} \gtrsim$ 10 GeV) is generally comparable to that over which the shock power peaks $t \sim t_{\rm pk}$ (see Eq.~\eqref{eq:Emax_Lgamma}).  Our fiducial model, properly calibrated using X-ray constraints on the size of the particle acceleration zone and assuming efficient magnetic field amplification (Sec.~\ref{sec:Bfield}), therefore explains why classical novae exhibit cut-offs in their $\gamma$-ray spectra above $\sim 1-10$ GeV near peak luminosity.  This contrasts with earlier work (e.g., \citealt{Metzger+16}), which neglected suppression of the particle acceleration layer due to turbulent mixing, and found that novae could sometimes reach TeV energies even at $t\sim t_{\rm pk}$.

\citet{Craig+25} found that classical novae with observed $\gamma$-ray emission exhibit a correlation (with large scatter) of the form $L_{\gamma} \approx 10^{35}\,{\rm erg\,s^{-1}}\,(v_{\rm sh}/10^{3}\,{\rm km\,s^{-1}})^{3},$ where $v_{\rm sh} \approx \Delta v$ is the shock velocity inferred from the difference of slow and fast outflow components in the optical spectra, $\Delta v \equiv v_{2}-v_{1}$.  Taking $v_{\rm sh} \approx v_{\rm f}/2$ for the toy model, we see that Eq.~\eqref{eq:Lgammamax} roughly matches the normalization found by \citet{Craig+25} for typical values of $M_{\rm env} \sim 10^{-5}-10^{-4}M_{\odot}$ and $\tau$.  For fixed $\tau$ and $M_{\rm env}$, the predicted dependence $L_{\gamma}^{\rm pk} \propto M_{\rm env}v_{\rm sh}^{2}/\tau$ on shock velocity (Eq.~\eqref{eq:Lgammamax}) is somewhat shallower than the scaling $\propto v_{\rm sh}^{3}$ found by \citet{Craig+25}.  However, faster novae (smaller $t_{\rm 2} \sim \tau$) exhibit higher ejecta velocities and are predicted to occur for lower envelope/ejecta masses; for example, taking the empirical relation $\tau \propto v_{\rm f}^{-2}$ found by \citet{McLaughlin60,Warner95}, our model would predict $L_{\gamma}^{\rm pk} \propto M_{\rm env} v_{\rm sh}^{4}$. 

\subsection{Delayed TeV Emission}

\label{sec:TeV}

Our example model in Figure \ref{fig:evolution} shows that shocks in classical novae can also power TeV emission, but only after a significant delay $t \gg \tau$, as the shock radius and speed grow in time (see also the discussion after Eq.~\eqref{eq:Emax}).  However, because the shock luminosity also becomes weaker at late times $t \gg \tau$, the maximum TeV luminosity will be appreciably lower than the maximum GeV luminosity.

Nevertheless, the right panel of Figure \ref{fig:luminosities} shows that a subset of the {\it LAT-}detected novae may also be observable with H.E.S.S. or other current-generation IACTs (e.g., MAGIC, VERITAS). In the case of TeV emission, the strong dependence of $E_{\rm max}$ on $v_{\rm f}$ implies that only novae with large outflow velocities are potentially detectable. Furthermore, ``fast" novae, with small $\tau$, $t_2$, tend to be optically thick to TeV $\gamma$-rays for most of their evolution, making detection challenging. 

 If a correlation between speed class and ejecta velocity of the form $v_{2} \propto v_{\rm f} \propto t_2^{-1/2} \sim \tau^{-1/2}$ \citep{McLaughlin60,Warner95} holds, then fast novae with large $\tau$ may be relatively rare. However, as we will show below, a number of real, \emph{Fermi} LAT detected novae met our criteria for TeV detectability. We argue that these novae were not detected because of the delay between $t_{\rm pk}$ and the time when a nova can first produce (unabsorbed) TeV $\gamma$-rays, $t_{\rm TeV,\gamma}$. In Figure \ref{fig:ttev}, we show $t_{\rm TeV,\gamma}$ (absolute value and normalized to $t_{\rm pk}$) as a function of $v_{\rm f}$ and $\tau$. For potentially TeV-detectable novae, the delay between $t_{\rm pk}$ and $t_{\rm TeV, \gamma}$ can be days or even weeks, with $t_{\rm TeV,\gamma} = t_{\rm pk}$ only for slow (large $\tau$) novae with extremely large (possibly unfeasible) $v_{\rm f}$.

\begin{figure}
    \centering
    \includegraphics[width=1.0\linewidth]{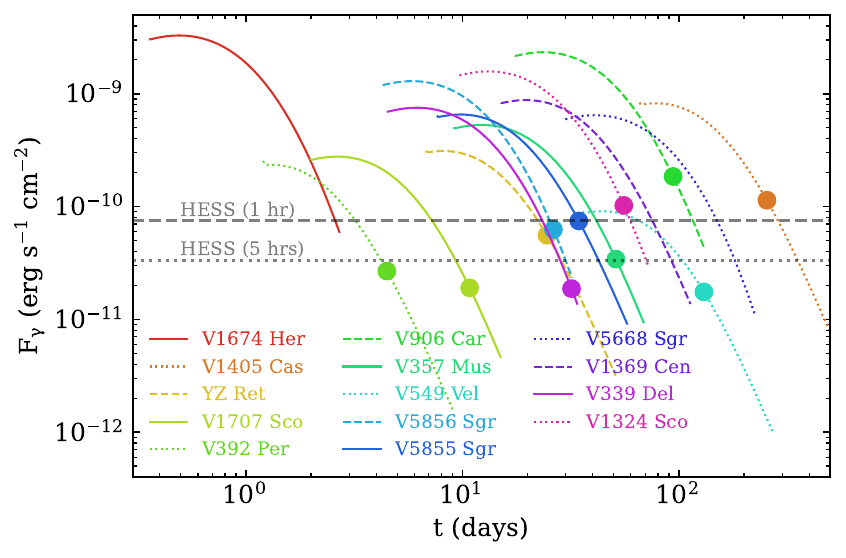}
    \caption{Predicted $\gamma$-ray light curves of the classical novae detected by \emph{Fermi} LAT as compiled by \citep{Craig+25}.  To assign the parameters of our model to each nova, we (1) relate the final wind speed ($v_{\rm f}$) to the ``fast" ejecta component observed spectroscopically ($v_{2}$) according to $v_{\rm f} = 2v_{2}$; (2) relate the envelope removal timescale ($\tau$) to the nova speed class according to $t_2 \approx 2\tau$; (3) chose an ejecta mass $M_{\rm env}$ necessary to reproduce the observed peak $\gamma$-ray luminosity. All other parameters are set to their fiducial values (Table~\ref{tab:modelparams}). The luminosity when the nova is first predicted to produce TeV $\gamma$-rays is denoted with a filled circle. H.E.S.S. $5\sigma$ detection thresholds are denoted with a gray dashed and dotted lines. Our model suggests that a subset of the $\gamma$-ray bright novae detected with \emph{Fermi} LAT may have also been detectable with current instruments, albeit days to weeks after the $\gamma$-ray peak.}
    \label{fig:novaobs}
\end{figure}

In light of this delay, we apply our model to the population of novae observed by \emph{Fermi} LAT to estimate the subset that may have been detectable with H.E.S.S. Namely, for each nova, we consider the observationally-constrained $\tau$ and $v_{\rm f}$ \citep[see Table 1 in][]{Craig+25}: we take $\tau =0.5t_2$ and $v_{\rm f} = 2v_2$, where $v_2 \approx v_{\rm sh}$ is the spectroscopically inferred velocity of the ``fast" component. We then set $M_{\rm env}$ by requiring that our model reproduces the observed $L^{\rm pk}_\gamma$ (namely, we invert Equation \ref{eq:Lgammamax}). Assuming our fiducial $\xi_{\rm CR}$, $\xi_{\rm B}$, $f_{\rm X}$, and $f_{\rm \Omega}$, we then estimate the evolution of $L_{\gamma}$ and $E_{\rm max}$; in Figure \ref{fig:novaobs}, we show the resulting $\gamma$-ray light curves and denote $L_{\gamma}$ when the maximum $\gamma$-ray energy first exceeds one TeV (overlaid circles). Filled circles further indicate that the nova is optically thin to TeV $\gamma$-rays. Of the 14 $\gamma$-ray detected novae, we predict that $\sim 5$ may have been detectable at TeV energies, albeit weeks after $t_{\rm pk}$.

We also note searches have been conducted in the past for TeV $\gamma$-rays from relatively fast novae, such as V339 Del \citep{Ahnen+15} and V392 Per \citep{Albert+22}, neither of which yielded a significant detection. However, in our model,  these novae are poor TeV candidates (both fall below H.E.S.S. limits; see Figure \ref{fig:novaobs}). Namely, our model prefers slower novae as potential TeVatrons, provided their shock velocities and/or $\gamma$-ray luminosities are sufficiently large.

\begin{figure*}
    \centering
    \includegraphics[width=0.5\linewidth, clip=true,trim= 0 0 0 0]{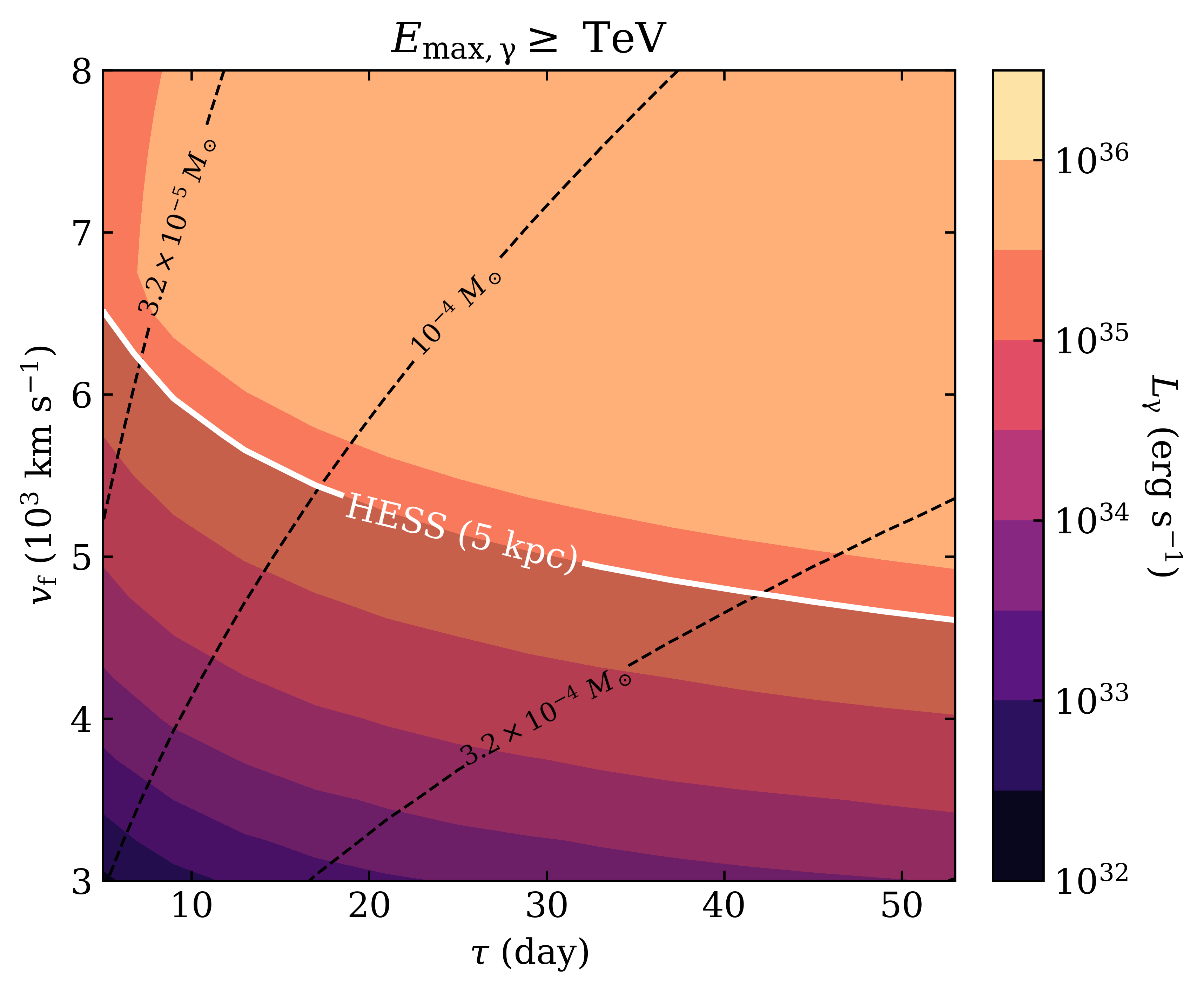}
    \includegraphics[width=0.49\linewidth, clip=true,trim= 0 0 0 0]{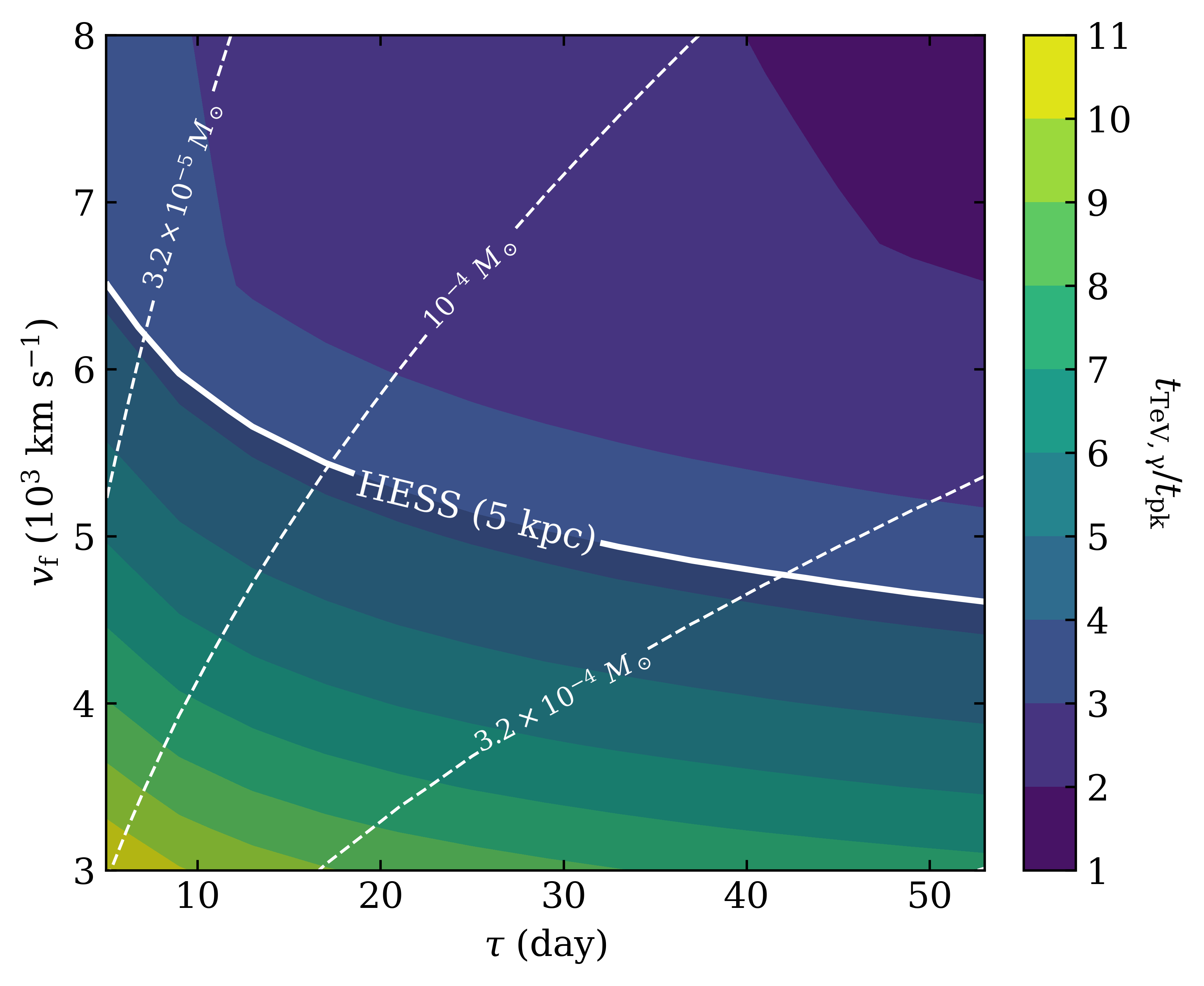}
    \caption{\emph{Left:} Same as Figure \ref{fig:luminosities} (right), except that the peak $\gamma$-ray luminosity ($L_\gamma^{\rm pk}$) is fixed to that of our fiducial nova, and $M_{\rm env}$ is varied accordingly (dashed contours). \emph{Right:} Same as Figure \ref{fig:ttev} (right), but again fixing $L_\gamma^{\rm pk} =2\times10^{36}$ erg s$^{-1}$. All other parameters are set according to our fiducial nova. }
    \label{fig:Lfixed}
\end{figure*}

We propose that the procedure outlined above can be applied to future nova outbursts in order to predict whether (and when) detectable TeV emission will occur (see Figure \ref{fig:Lfixed}, which shows $L_{\gamma}(t_{\rm TeV,\gamma})$ and $t_{\rm TeV,\gamma}/t_{\rm pk}$ as a function of $v_{\rm f}$ and $\tau$, assuming a fixed $L_{\gamma}^{\rm pk}$). Namely, early observations (around $t_{\rm pk}$) can be used to constrain $\tau$, $v_{\rm f}$, and $M_{\rm env}$. Depending on the available multi-wavelength observations, a more refined analysis might also use X-ray and optical fluxes to constrain $f_{\rm X}$, radio morphology to constrain $f_{\Omega}$, and/or the early-time $\gamma$-ray cutoff to constrain $\xi_{\rm B}$. This leaves only $\Delta_{\rm s}/R_{\rm s}$ as a free parameter, though it is reasonably well-constrained by theoretical arguments (see Section \ref{subsec:hydro}).

\section{Discussion}
\label{sec:discussion}
Herein we discuss additional implications, limitations, and potential extensions of our model.

\subsection{Time Evolution of the X-ray Efficiency?}
\label{sec:fX}

Our fiducial calculations assume a constant X-ray efficiency 
$f_X \equiv L_X/L_{\rm sh}$, motivated by the values inferred at early 
times when the GeV emission is detected and contemporaneous X-ray 
observations are available. However, $f_X$ may itself evolve as the shock velocity and ambient density change. Because the thickness of the hot post-shock layer scales as $\Delta_{\rm down} \propto f_X \Delta_{\rm rad}$ 
(Eq.~\eqref{eq:deltaDS}), and the maximum proton energy satisfies $E_{\max} \propto \Delta_{\rm down}$ (Eq.~\eqref{eq:Emaxcondition}), any systematic variation in 
$f_X$ will directly modify the predicted evolution of $E_{\max}$.   If $f_X \propto 
v_{\rm sh}^{\alpha}$ with $\alpha < 0$, then combining this scaling with 
Eq.~\eqref{eq:Emax} gives 
\begin{equation}
E_{\max} \propto f_X v_{\rm sh}^{9/2} 
\propto v_{\rm sh}^{9/2+\alpha}.
\label{eq:Emax_scaling}
\end{equation}

On the one hand, higher shock velocities may drive more vigorous turbulence through Rayleigh-Taylor and thin-shell instabilities and related mixing 
processes, leading to more efficient entrainment of hot gas into the 
cool dense shell \citep{Metzger+25}. In this case the X-ray emitting layer would be further 
suppressed at large $v_{\rm sh}$, reducing $f_X$. Indeed, the analytic 
estimate of \cite{Metzger+25} implies a minimum efficiency 
$f_{X,\min} \propto v_{\rm sh}^{-3}$ (Eq.~\eqref{eq:fXmin}), suggesting that stronger 
shocks could exhibit systematically smaller $f_X$. Taking $\alpha = -3$ in Eq.~\eqref{eq:Emax_scaling} would reduce the velocity dependence $E_{\max} \propto v_{\rm sh}^{3/2}$, significantly slowing the growth 
of the maximum particle energy relative to our fiducial model.

On the other hand, as the wind density declines at late times and the 
shock radius expands, the post-shock cooling time increases and the 
shock may become less strongly radiative. In this regime mixing could 
become less efficient, allowing a thicker hot layer and a larger $f_X$. 
If instead $f_X$ increases with velocity (i.e., $\alpha > 0$), then 
$E_{\max}$ would grow even more rapidly than in our baseline model, 
potentially advancing the onset of TeV emission.

Thus, while the qualitative prediction that $E_{\rm max}$ increases with 
time is robust, the 
precise rate at which $E_{\max}$ rises, and hence the timing and peak 
luminosity of any TeV component, depend on the poorly constrained 
evolution of $f_X$. Improved constraints on the velocity dependence of 
the X-ray efficiency from coordinated X-ray and $\gamma$-ray 
observations for particularly bright events will therefore be essential to refine predictions for the TeV emission from classical novae.

\subsection{Magnetic Field Amplification}
\label{sec:Bfield}

Throughout our analysis, we have assumed that a constant fraction of the ram pressure at the shock is converted into magnetic pressure: $\xi_{\rm B} \equiv P_{\rm B}/(\rho_{\rm sh}v_{\rm sh}^2) = const$. The white dwarf wind may also contribute to the magnetic field at the shock; taking the field on the white dwarf surface (with radius $R_{\rm WD} \sim 1000$ km) to be $\sim 10^5$ G and assuming a steady wind such that $B\propto1/R_{\rm s}$, one obtains $B_{\rm down}\sim1$ G at $t_{\rm pk}$ for our fiducial model. This field is comparable to that obtained by assuming magnetic field amplification with efficiency $\xi_{\rm B} = 0.01$. However, in practice, the burning layer comprising the ejecta is highly turbulent, such that $B\propto 1/R_{\rm s}$ is very optimistic. If, instead, the field behaves as an adiabatic gas with index $\gamma_{\rm ad}=4/3$ such that $B^2\propto \rho^{\gamma_{\rm ad}}$, we obtain $B\propto R_{\rm s}^{-4/3}$. The resulting $B_{\rm down}$ is negligible with respect to the amplified one, and insufficient to produce GeV $\gamma$-rays by $t_{\rm pk}$.

That being said, a small subset ($\lesssim 10\%$) of white dwarfs may be ``magnetic", with surface fields approaching $10^6-10^7$ G \citep{Ferrario+15}. 
In the most extreme case ($B\sim10^7$ G), such a surface field leads to $B_{\rm down}$ that is comparable to the amplified field in our fiducial model, even accounting for a turbulent ejecta ($B\propto R_{\rm s}^{-4/3}$). Thus, there may be a small subset of classical novae capable of achieving even higher energy protons (and thus $\gamma$-rays).

Still, in all but the most extreme (and likely rare) cases, reproducing observations of GeV $\gamma$-rays by $t\approx t_{\rm pk}$ \emph{requires} the amplification of turbulent magnetic fields ($\xi_{\rm B}\gtrsim 0.01$) with significant power on the gyroresonant scale (in order to have Bohm diffusion). Thus far, however, we have remained agnostic to the microphysical processes responsible. 

In supernova remnants (SNRs), magnetic field amplification is thought to occur via the non-resonant hybrid (``Bell") instability \citep{Bell04}, in which escaping protons drive strong fluctuations with $\delta B/B\gg 1$. This instability saturates when approximate equipartition is reached between the magnetic pressure and the escaping current, such that $\xi_{\rm {B}} \sim v_{\rm sh}\xi_{\rm CR}$. As such, Bell amplification implies an even stronger dependence of $E_{\rm max}$ on $v_{\rm sh}$ than that proposed in Eqs.~\eqref{eq:Emax} and \eqref{eq:Emax_Lgamma}, potentially leading to TeV emission at slightly earlier times. However, whether Bell operates efficiently at reverse shocks remains an open question in the literature.

At the same time, the highly turbulent environment coupled with the strong cosmic ray pressure gradient around the shock may contribute to magnetic field amplification via the acoustic instability and/or turbulent dynamo \citep{Beresnyak+09, drury+12}. In both instabilities, pre-existing magnetic fluctuations become highly amplified, with saturated fields comparable to those generated via the Bell instability. The fact that X-ray suppression via mixing (as discussed in \citealt{Metzger+25}) requires strong turbulence on scales from $\Delta_{\rm down} $ down to $\sim 10^{-4}\Delta_{\rm down}$ makes this scenario particularly promising, as it implies magnetic field amplification on gyroresonant scales (recall that, with Bohm diffusion, $r_{\rm L}(E_{\rm max}) = 3\Delta_{\rm down}v_{\rm sh}/c$).

More broadly, $\gamma$-ray observations of classical novae can shed light onto the nature of magnetic field amplification at astrophysical shocks. In particular, the time evolution of $E_{\rm max}$ constrains that of $\xi_{\rm B}$, which in turn encodes information about the microphysical processes at play. In this regard, and given that classical novae may produce TeV emission at late times, observations with upcoming IACTs such as the Cherenkov Telescope Array \citep[CTA,][]{CTA} may be especially useful; our model suggests that instruments such as CTA will have sufficient sensitivity not only to detect TeV $\gamma$-rays from novae but also to measure their evolution.

\subsection{Radio Emission and the Forward Shock}

In general, relativistic electrons accelerated at nova shocks, or secondary $e^\pm$ pairs produced through charged pion decay following proton-proton collisions, are also expected to power synchrotron emission at radio frequencies (e.g., \citealt{Taylor+87,Vlasov+16}). However, such synchrotron emission from the powerful reverse shock is not directly observable at early times because it is strongly absorbed by the dense, radiatively cooled shell. Within our toy model, the free--free optical depth through the cool shell is 
\be
\begin{split}
\tau_{\rm ff} &= \alpha_{\rm ff}\Delta_{\rm s} \\
\underset{t \gg \tau}&\approx 1.4\times 10^{8}\frac{M_{\rm env,-4}^{2}}{\tau_{20}^{5}}\left(\frac{v_{\rm f, 8}}{2}\right) ^{-5}\left(\frac{\nu}{10\,{\rm GHz}}\right)^{-2} \\ &\times T_{\rm s,4}^{-3/2}\left(\frac{\Delta_{\rm s}}{10^{-2}R_{\rm s}}\right)^{-2}\left(\frac{t}{\tau}\right)^{-5},
\end{split}
\ee
where $\alpha_{\rm ff} = \alpha_0 \nu^{-2} T_{\rm s}^{-3/2} n_{\rm s}^{2}$ is the free-free absorption coefficient and we take $\alpha_0 \approx 0.1$ cm$^{5}$K$^{3/2}$s$^{-2}$ \citep{Rybicki&Lightman79}.  Even if the shell is only partially ionized, the large density implies that radio emission from the reverse shock is heavily attenuated, with $\tau_{\rm ff} \gg 1$ at GHz frequencies near peak shock power. Thus, although the reverse shock dominates the optical reprocessing and $\gamma$-ray emission in the calorimetric regime, it does not contribute appreciably to the observed radio flux at early times.

There are, however, two natural channels by which synchrotron emission can escape.  First, it could originate from the forward shock generated as the dense shell collides with ambient matter surrounding the white dwarf on larger scales (e.g., \citealt{Vlasov+16}).  In the internal shock framework adopted here, the forward shock is generally weaker than the reverse shock. Because we assume a wind with a monotonically increasing velocity (Eq.~\eqref{eq:vw}), most of the kinetic power is dissipated at the reverse shock, where the fast wind collides with the previously swept-up slow material (Fig.~\ref{fig:diagram}). However, while our toy model assumption of a smoothly accelerating wind is a convenient parameterization, real nova outflows likely exhibit a broader distribution of velocities, particularly during the earliest phases of the thermonuclear runaway (e.g., \citealt{Starrfield+98,Chomiuk+21}).  As a result, the dense shell may be propagating into a low density medium but not into a vacuum.   Although the forward shock is not expected to contribute as significantly to the optical or $\gamma$-ray luminosity, its radio emission may dominate because it doesn't need to propagate through the dense shell.  

A second source of synchrotron emission can arise if relativistic pairs leak upstream of the shocks and radiate in the unshocked ejecta or circumstellar medium. Such escape could occur once the shock weakens or if magnetic turbulence permits diffusion on scales comparable to the shock radius. In the symbiotic nova V3890 Sgr, \citet{Molina+26} show that escaping non-thermal particles can illuminate extended regions of the red giant wind, explaining a diffuse synchrotron halo seen in this event. A similar, though likely less dramatic, effect may operate in classical novae, particularly at late times when the column density through the shell declines.

\subsection{Implications for Symbiotic Novae}

Symbiotic novae differ from classical novae in that the external medium into which the ejecta expand is provided by the dense wind of a red giant companion rather than by slower ejecta released earlier in the outburst. In classical novae, collisions between fast and slow white dwarf ejecta produce a dense, radiatively cooled shell. Turbulent mixing between the hot post-shock gas and this shell suppresses the X-ray–emitting layer, limiting the radial thickness of the particle acceleration region and thereby constraining the maximum proton energy. In symbiotic systems, however, the circumstellar environment is established prior to eruption and is typically described by a red giant wind with density profile $\rho \propto r^{-2}$ \citep{Seaquist+90,Seaquist+93}.

At small radii the wind density can be sufficiently high for the forward shock driven into the wind to become radiative, potentially forming a cooled shell analogous to that in classical novae. Early hard X-ray emission in symbiotic novae such as RS~Oph and V407~Cyg indeed indicates strong shocks propagating into dense pre-existing material \citep{Sokoloski+06,Abdo+10,Page+15}. However, because the wind density declines with radius, the shock rapidly transitions from radiative to partially or fully adiabatic behavior. In the absence of a massive cool shell, the post-shock region may therefore be substantially thicker than in the mixing-suppressed classical nova scenario discussed above, potentially allowing larger acceleration zones and higher maximum particle energies at early times.

High-resolution radio imaging shows that the circumstellar environments of symbiotic novae are often highly asymmetric. In V3890~Sgr, VLBI observations reveal bipolar expansion \citep{Molina+26}, possibly shaped by an equatorial density enhancement (EDE) in the orbital plane \citep{Orlando+17}. Similar structures have been inferred in RS~Oph \citep{Munari+22,Lico+24} and are expected theoretically from gravitational focusing and wind Roche-lobe overflow \citep{Mohamed+12,Walder+08}. In such geometries the nova ejecta interact simultaneously with a lower-density polar wind and a denser equatorial component. As suggested for V3890~Sgr \citep{Molina+26}, the radio-emitting shocks may trace the lower-density polar regions, while the $\gamma$-rays originate in the denser equatorial material where proton–proton losses are more efficient \citep{Diesing+23}.

This configuration is qualitatively analogous to the internal-shock geometry in classical novae, with the EDE effectively playing the role of the slow ejecta. If the interaction with the equatorial overdensity is radiative and subject to turbulent mixing, a thin post-shock layer may again form, constraining the acceleration region thickness. If mixing is weaker or the shock becomes adiabatic more rapidly, however, the acceleration region may be significantly thicker. In either case, a robust prediction is that as the shock expands and the ambient density declines, the effective size of the acceleration zone grows and the maximum particle energy should increase with time, at least until the large swept-up mass begins to decelerate the shock.

This behavior naturally leads to delayed TeV emission in symbiotic novae. TeV $\gamma$-rays have now been detected from RS~Oph during its 2021 eruption by H.E.S.S. \citep{Acciari+22}, following contemporaneous GeV emission observed by \emph{Fermi} \citep{Cheung+22b}. The larger binary separations and extended circumstellar environments of symbiotic systems imply larger characteristic shock radii and longer dynamical times than in classical novae. Consequently, the maximum particle energy may reach the TeV range once the shock has propagated to sufficiently large radii and the system becomes optically thin to $\gamma$–$\gamma$ absorption. The delayed onset of TeV emission relative to the GeV peak observed in RS~Oph is qualitatively consistent with this picture.

An additional difference between classical and symbiotic novae may lie in the ambient magnetic field. In classical novae, magnetic fields originating on the white dwarf are strongly diluted by expansion, requiring efficient local amplification (parameterized by $\xi_B$) to achieve Bohm-like diffusion and GeV emission. In symbiotic systems, however, the red giant wind may carry a stronger large-scale field. For plausible red giant surface fields of order $\sim$G and flux freezing in the outflow, magnetic field strengths of $B \sim 10^{-3}$–$10^{-2}$~G at radii $\sim 10^{15}$–$10^{16}$~cm are reasonable \citep{Molina+26}. Such fields provide a larger seed for shock amplification and may facilitate acceleration to TeV energies \citep{Diesing+23}.


\section{Conclusion}
\label{sec:conclusion}
We developed a simple, parameterized toy model for internal shock interaction in classical novae and explored its implications for GeV–TeV $\gamma$-ray emission. In this framework, a fast wind collides with a shell comprised of the earlier, slower ejecta, generating radiative shocks that accelerate relativistic protons at the reverse shock. The model provides convenient analytic expressions for the time evolution of the shock/shell velocity (Eqs.~\eqref{eq:vs}, \eqref{eq:vsh}), radius (Eq.~\eqref{eq:Rs}), kinetic power (Eq.~\eqref{eq:Lsh}), immediate post-shock temperature and density (Eqs.~\eqref{eq:Tsh}, \eqref{eq:rhosh}), density and X-ray column through the dense shell (Eqs.~\eqref{eq:rhos}, \eqref{eq:NH}), as well as the luminosity (Eq.~\eqref{eq:Lgamma}) and maximum energy (Eq.~\eqref{eq:Emax}) of the gamma-ray emission.  The model framework may be expanded in future work to allow for more complex mass-loss history from the white dwarf, for example by allowing for temporal fluctuations in the wind velocity (which can give rise to multiple internal shocks and associated cool dense shells that eventually coalesce into one; e.g., \citealt{Steinberg&Metzger20}).   

For typical nova parameters, the high densities in the cool shell place the system in the calorimetric limit near the peak shock power, such that the $\gamma$-ray luminosity directly tracks the instantaneous shock power. This naturally accounts for the comparable durations of the optical and GeV emission and implies cosmic-ray acceleration efficiencies of order $\xi_{\rm CR} \sim 0.01–0.1$.  The toy model also connects nova observable, broadly consistent with observed correlations between ejecta velocity and gamma-ray luminosity \citep{Craig+25}. 

Small observed X-ray luminosities require strong suppression of the size of the hot post-shock region, which in turn limits the maximum particle energy. Near optical maximum, our model predicts proton energies of order $\sim 10$ GeV, consistent with the spectral cutoffs observed by \emph{Fermi} LAT.  The predicted spectral cut-off is generally smaller than previous predictions by \citet{Metzger+16}, who considered that photo-ionization rather than turbulent mixing sets the width of the acceleration region behind the shock.
Nevertheless, as the shock expands and the mass-loss rate declines, the maximum particle energy increases rapidly, potentially reaching $\gtrsim$ TeV on timescales of several $\tau$.

This evolution leads to a clear observational prediction: TeV emission, if present, should be delayed relative to the GeV peak and occur at lower luminosity. Detectability therefore depends not only on peak $\gamma$-ray luminosity, but also on ejecta velocity, characteristic timescale $\tau$ (or $t_2$), and the time required for the ejecta to become optically thin to gamma-gamma absorption. Applying this framework to the population of \emph{Fermi} detected novae suggests that a small subset of nearby, high-velocity systems could produce detectable TeV emission days to weeks after optical maximum.

Early-time optical and GeV observations can be used to estimate shock parameters $\tau$, $v_{\rm f}$, and $M_{\rm env}$, enabling predictions of the timing and brightness of any TeV component. Coordinated follow-up of fast, nearby novae at late times therefore provides a direct test of our model and offers constraints on magnetic field amplification and particle acceleration at radiative shocks.

\section*{Acknowledgments}

We thank Laura Chomiuk, Peter Craig, and Elias Aydi for their helpful comments. This work was supported in part by NASA (grants  80NSSC22K0807, 80NSSC24K0408, 80NSSC26K0300) and by Columbia University through the Research Stabilization Fund.  The Flatiron Institute is supported by the Simons Foundation.

\bibliographystyle{aasjournal}

\end{document}